\documentclass[onecolumn, aps,prx, showpacs, superscriptaddress, floatfix,nofootinbib]{revtex4-2}
\usepackage[T1]{fontenc}
\usepackage[utf8]{inputenc}
\usepackage{mathtools}
\usepackage{graphicx}
\usepackage{caption}
\usepackage{subcaption}
\usepackage{xcolor}
\usepackage{soul}
\usepackage{amsmath,latexsym}
\usepackage{array}
\usepackage{tikz,hyperref}
\usepackage{orcidlink}
\usepackage{float}
\usepackage{cleveref}
\usepackage[percent]{overpic}

\newcommand{\msun}{$M_\odot$}
\newcommand{\ft}[0]{\footnotesize}
\newcommand{\physrep}{Phys.~Rep.}
\newcommand{\mnras}{Mon.~Not.~Roy.~Astron.~Soc.}
\newcommand{\apjl}{Astrophys.~J.~Lett.}
\newcommand{\pasp}{PASP}
\newcommand{\aapr}{Astron.~Astrophys.~Rev.}
\newcommand{\araa}{Annu.~Rev.~Astron.~Astrophys.}

\newcommand{\jcap}{JCAP}

\begin{document}

\title{Hyperons and $\Delta$'s in rotating protoneutron stars: Local properties} 

\author{Franciele M. da Silva \orcidlink{0000-0003-2568-2901}} 
\email{franciele.m.s@ufsc.br}

\affiliation{Departamento de F\'isica, CFM - Universidade Federal de Santa Catarina, \\ Caixa Postal 5064, CEP 880.35-972, Florian\'opolis, SC, Brazil.}
\affiliation{Theoretical Astrophysics, Institute for Astronomy and Astrophysics, University of T\"{u}bingen, 72076 T\"{u}bingen, Germany}

\author{Adamu Issifu \orcidlink{0000-0002-2843-835X}} 
\email{ai@academico.ufpb.br}
\affiliation{Departamento de F\'isica, Instituto Tecnol\'ogico de Aeron\'autica, DCTA, 12228-900, S\~ao Jos\'e dos Campos, SP, Brazil} 
\affiliation{Laborat\'orio de Computa\c c\~ao Cient\'ifica Avan\c cada e Modelamento (Lab-CCAM)}

\author{Luis C. N. Santos \orcidlink{0000-0002-6129-1820}}
\email{luis.santos@ufsc.br}

\affiliation{Departamento de F\'isica, CFM - Universidade Federal de Santa Catarina, \\ Caixa Postal 5064, CEP 880.35-972, Florian\'opolis, SC, Brazil.}

\author{Tobias Frederico \orcidlink{0000-0002-5497-5490}} 
\email{tobias@ita.br}

\affiliation{Departamento de F\'isica, Instituto Tecnol\'ogico de Aeron\'autica, DCTA, 12228-900, S\~ao Jos\'e dos Campos, SP, Brazil} 
\affiliation{Laborat\'orio de Computa\c c\~ao Cient\'ifica Avan\c cada e Modelamento (Lab-CCAM)}

\author{D\'ebora P. Menezes \orcidlink{0000-0003-0730-6689}}
\email{debora.p.m@ufsc.br}

\affiliation{Departamento de F\'isica, CFM - Universidade Federal de Santa Catarina, \\ Caixa Postal 5064, CEP 880.35-972, Florian\'opolis, SC, Brazil.}

\begin{abstract}
  The structural evolution of rotating protoneutron stars encodes essential information about their observable signatures, while microscopic properties provide complementary knowledge to advance observational investigations. Using a relativistic mean-field model with density-dependent couplings that account for temperature and particle composition, we investigate rotation, neutrino-emission-driven changes in angular momentum, particle distributions, temperature profiles, and sound speed to probe the internal dynamics of protoneutron star matter. Additionally, we track the evolution of macroscopic quantities such as energy distribution and gravitational mass and establish direct links between microphysics and global evolution. Extending the framework of \textcolor{blue}{Phys.~Rev.~\textbf{D~112},~023007~(2025)}, which focuses on the global properties of rotating protoneutron star evolution, our results reveal that protoneutron star deformation and thermal evolution are governed by angular momentum, mass, and composition. Exotic matter (hyperons and $\Delta$-resonances) and rapid rotation enhance deformation leading to a reduction in core temperature, whereas slowly rotating stars like PSR J0740$+$6620 remain nearly spherical. Our predicted equatorial radii for PSR J0740$+$6620, $13.0\ \mathrm{km} < R_e < 13.5\ \mathrm{km}$, are consistent with NICER measurements. These findings constrain the EoS, requiring a self-consistent treatment of rotation, mass-dependent compression, and composition-driven modeling to accurately model protoneutron star evolution in the context of multi-messenger astrophysics.
\end{abstract}

\maketitle

\section{Introduction}

In scenarios where the collapse of a massive stellar core does not lead to black hole formation, the remnant can become a hot protoneutron star (PNS), potentially exhibiting rapid rotation at the early stages after supernova explosion. During the early evolution, the PNS exhibits strong radial gradients of thermodynamic quantities. Local properties such as the energy density, pressure, temperature, and lepton fraction evaluated at some radius play an important role in determining the internal dynamics and stability of the star \cite{Janka:2006fh}. Following the supernova, the compact remnant is initially hot and rich in leptons \cite{Prakash:1996xs}. Neutrinos begin to escape through diffusion, gradually lowering the lepton number and contributing to internal heating \cite{Burrows:2020qrp}. Over time, this leads to cooling, and the object evolves into a cold-catalyzed neutron star (NS) with low neutrino content, a process that occurs on the order of tens of seconds, followed by much longer-term cooling over thousands to millions of years~\cite{Pons:1998mm, burrows1986}.

During the evolution of a PNS, rotation is of central importance in the evolution and structure of the compact remnant. In principle, during a core-collapse supernova, the conservation of angular momentum of the progenitor star can lead to the formation of a newly born PNS that is rapidly rotating. Observations indicate that stars undergoing supernova collapse often possess high rotational velocities in their final stages \cite{vanbelle2012A&ARv,fukuda1982PASP}.
In particular, the Keplerian limit defines the maximum rotation rate a NS can sustain before mass loss begins at the equator due to centrifugal forces overcoming gravity. This critical frequency depends on the star's mass and internal structure and marks a boundary between stable rotation and dynamical instability. In this context, developments in the field point to a connection between long-duration gamma-ray bursts and the formation of magnetars spinning near the Keplerian rotation limit, or with early stages leading to the collapse into a rapidly rotating black hole~\cite{langer2012ARA&A, woosley2006ARA&A}.

This work extends the framework developed in \cite{daSilva:2025cfe}, advancing the study of PNS evolution by incorporating rotational effects in hypernuclear matter composed of nucleons, hyperons ($\Lambda$, $\Sigma$, and $\Xi$)~\cite{Weissenborn:2011kb, Issifu:2024htq, thakur2025radial}, and $\Delta$-resonances~\cite{Issifu:2023qyi}. PNSs, formed through core-collapse supernova \cite{Lattimer:2000nx, langer2012presupernova}, exist under extreme conditions \cite{Glendenning2000} where the baryon composition crucially influences the early cooling phase, lasting approximately 30 seconds \cite{Pons:1998mm}, while rotation profiles significantly affect the star's maximum mass and stability \cite{Stergioulas:2003yp}. The emergence of exotic degrees of freedom, such as hyperons and $\Delta$-resonances \cite{Issifu:2023qyi}, softens the equation of state (EoS), leading to greater compactness and, consequently, influencing rotational dynamics and the star's internal structure. We can find some works in the literature that have investigated the influence of $\Delta$-resonances in the evolution of PNSs~\cite{prakash1992ApJ,Dexheimer:2008ax,raduta2022EPJA,sedrakian2022,Issifu:2023qyi}. 
Our extended model addresses several key gaps in the cited work and the broader literature. {More importantly, we consider a star with a fixed baryon mass, allowing us to disentangle changes in the stellar properties that arise solely from thermal evolution and rotational dynamics.} Specifically, the work investigates the hole rotation and angular momentum loss by neutrino emission during the Kelvin-Helmholtz evolutionary stage of a rotating PNS, considering multi-component matter. We analyze properties within the stellar matter, such as the spatial distribution of particle species, temperature, and the sound speed during the evolution of the PNS. In addition, we investigate how the rotation and composition of the PNS can influence properties such as the degree of deformation, the moment of inertia, and energy distribution as the star evolves. By integrating these aspects, our approach aims to provide a more comprehensive understanding of PNS dynamics under controlled astrophysical conditions. 

This work is organized as follows: in \Cref{sec_model} we concisely present the methodology utilized in this work, specifically, in \Cref{sec_eos} we briefly present the EoS models we are considering, and in \Cref{sec_prop} we explain how we calculate some of the PNS properties under analysis. In \Cref{sec_160} we present our results for a hypothetical PNS that is rapidly rotating and has a baryonic mass of $M_0=1.60$~\msun~and in \Cref{sec_nicer} we study a more realistic case of a PNS that has the same gravitational mass and frequency of PSR J0740$+$6620 when the star becomes a cold catalyzed NS. Finally, in \Cref{remarks}, we present our main conclusions.

\section{Model Formalism} \label{sec_model}

In this work, we investigate the local and also some global properties of individual PNSs as they evolve along four snapshots. The snapshots we consider follow the same sequences of the earlier study published in Ref.~\cite{daSilva:2025cfe}, therefore, the first snapshot represents an initial stage of evolution when the star is neutrino opaque and lepton rich with low entropy per baryon $(s_B=1, \,Y_L=0.4)$, the second snapshot represents the stage in which the PNS is deleptonizing through neutrino diffusion, this reduces the lepton fraction in the stellar matter and the entropy per baryon increases due to shock heating $(s_B=2, \,Y_L=0.2)$, the third snapshot shows the evolutionary stage when the PNS becomes neutrino transparent but still has a high entropy per baryon $(s_B=2, \,Y_{\nu_e}=0)$, finally, the forth snapshot represents the last stage of evolution of the PNS when it becomes a cold and catalized NS $(T=0\, {\rm MeV})$. We also consider three possible compositions for the matter inside the PNS, that is, leptons and nucleons $(N)$, leptons plus nucleons plus hyperons $(NH)$, and leptons plus nucleons plus hyperons plus delta resonances $(NHD)$. Below, we briefly present the EoSs used in this work and how we calculate some of the stellar properties we are going to analyze.

\subsection{Governing EoS} \label{sec_eos}

The interactions among baryons are modeled within the framework of quantum hadrodynamics (QHD)~\cite{Serot:1997xg}, where the baryon-baryon forces are mediated by the exchange of massive mesons \(i = \sigma, \omega, \rho, \phi\). Here, \(\sigma\) is a scalar meson, \(\omega\) and \(\phi\) are vector-isoscalar mesons, with \(\phi\) carrying hidden strangeness, and \(\rho\) is a vector-isovector meson. The Lagrangian density governing the system is given by
\begin{equation}
    \mathcal{L} = \mathcal{L}_{\rm H} + \mathcal{L}_\Delta + \mathcal{L}_{\rm m} + \mathcal{L}_{\rm L},
\end{equation}
where \(\mathcal{L}_{\rm H}\) describes the baryon octet, \(\mathcal{L}_\Delta\) the $\Delta$-resonances, \(\mathcal{L}_{\rm m}\) the meson fields, and \(\mathcal{L}_{\rm L}\) the free leptons. The explicit forms are:
\begin{align}
    \mathcal{L}_{\rm H} &= \sum_{b} \bar{\psi}_b \left[ i\gamma^\mu \partial_\mu 
    - \gamma^0 \left( g_{\omega b} \omega_0 + g_{\phi b} \phi_0 + g_{\rho b} I_{3b} \rho_{03} \right)
    - \left( m_b - g_{\sigma b} \sigma_0 \right) \right] \psi_b, \label{h}\\
    \mathcal{L}_{\Delta} &= \sum_{d} \bar{\psi}_d \left[ i\gamma^\mu \partial_\mu 
    - \gamma^0 \left( g_{\omega d} \omega_0 + g_{\rho d} I_{3d} \rho_{03} \right)
    - \left( m_d - g_{\sigma d} \sigma_0 \right) \right] \psi_d, \label{d}\\
    \mathcal{L}_{\rm m} &= -\frac{1}{2} m_\sigma^2 \sigma_0^2 
    + \frac{1}{2} m_\omega^2 \omega_0^2 
    + \frac{1}{2} m_\phi^2 \phi_0^2 
    + \frac{1}{2} m_\rho^2 \rho_{03}^2, \label{m}\\
    \mathcal{L}_{\rm L} &= \sum_L \bar{\psi}_L \left( i\gamma^\mu \partial_\mu - m_L \right) \psi_L. \label{l}
\end{align}
The $\Delta$ resonances are modeled using a Rarita–Schwinger Lagrangian, $\mathcal{L}_{\Delta}$, which includes a vector-valued spin projection absent in the baryon octet, represented by Dirac spin-$\tfrac{1}{2}$ fields (${\psi}_b$). However, within the mean-field approximation, the spin-$\tfrac{3}{2}$ $\Delta$ particles satisfy equations of motion analogous to those of the spin-$\tfrac{1}{2}$ baryons \cite{dePaoli:2012eq}. Free leptons are included to ensure charge neutrality of the system. The EoS is derived by computing the energy-momentum tensor from the full Lagrangian density \(\mathcal{L}\), which encapsulates the microphysical interactions. This formalism builds on the framework established in~\cite{daSilva:2025cfe} (the current work extends it for a comprehensive study), where detailed derivations and coupling prescriptions are provided. Further discussion of the EoS and its constraints can be found in~\cite{Lopes:2022vjx, Issifu:2023qyi, Issifu:2024fuw, Issifu:2024htq}. All expressions are given in natural units where $\hbar = c = G =k_B= 1.$

\subsection{Rotational Properties} \label{sec_prop}

The rotational treatment follows the formalism developed in~\cite{daSilva:2025cfe}, which concentrated on global stellar properties. 
The spacetime metric describing a star in rotational equilibrium can be assumed to be stationary and axially symmetric, that is, it can be given in quasi-isotropic coordinates $t$, $r$, $\theta$, $\phi$ by:
\begin{equation}
    ds^2 = -e^{2 \nu} dt^2 + e^{2 \alpha} \left( dr^2 + r^2 d \theta^2 \right) + e^{2 \beta} r^2 \sin^2{(\theta)} \left( d \phi - \omega dt \right)^2,
\end{equation}
where $\nu$, $\alpha$, $\beta$ and $\omega$ are the metric functions, which depend on $r$ and $\theta$. The equatorial radius $R_e$ of the star is given by the circumference in the equatorial plane, where the pressure vanishes, divided by $2 \pi$
\begin{equation}
    R_e = r e^{\beta} |_{r=r_e,\, \theta=\pi/2}.
\end{equation}
Here, we extend that framework to explore local thermodynamic and rotational impacts on the structural evolution of the PNSs. We extract the polar radius \( R_p \) and equatorial radius \( R_e \) from the global structure of the star and quantify the degree of rotational deformation, i.e., the deviation from spherical symmetry, using the percentage deformability defined as \cite{1994ApJ...424..823C}:
\begin{equation}
    f_R = \dfrac{R_e - R_p}{R_e} \times 100\%.
\end{equation}
In this framework, static NSs correspond to the spherical limit where  $R_p = R_e$, implying $f_R = 0\%$. This configuration serves as the baseline for quantifying the degree of rotational deformation. 

When a neutrino is emitted from the surface of a rotating PNS in a given direction, it carries a four-momentum $p^{\alpha}$. A local comoving observer, characterized by a four-velocity $u^{\alpha}$, measures the energy carried away by the neutrino as \cite{Baumgarte1998ApJ}
\begin{equation}
    \delta W^{\nu_e} = -u^{\alpha}p_{\alpha} = -u^{t}p_{t} -u^{\phi}p_{\phi}
\end{equation}
In this context, the component $p_t$ is associated with the change in gravitational mass $dM$, while $p_{\phi}$ is related to the change in angular momentum $dJ$, such that \cite{Zeldovich1971reas.book}:
\begin{equation}
    dM = \Omega dJ - \frac{\delta W^{\nu_e}}{u^t}, \label{eqdw} 
\end{equation}
where $\Omega = u^{\phi}/u^{t}$ is the angular velocity of the star measured by an observer at infinity. Starting from Eq.~(\ref{eqdw}) and doing some considerations, it is possible to arrive to the following empirical relation to describe the angular momentum loss due to neutrino emission~\cite{janka2004,martinon2014PhRvD}:
\begin{equation}
    \frac{\dot{J}}{J}=q \frac{f(\lambda)}{k_n} \frac{\dot{M}}{M}, \label{eqj1}
\end{equation}
where $q$ is an efficiency factor: $q = 0$ for central emission, $q = 1$ for uniform emission without scattering, and $q = \frac{5}{3}$ for diffusion. The function $f(\lambda)$ is defined by $f(\lambda) = 0.60\, (1 + 4\lambda)/(1 + 2\lambda)$, and accounts for stellar deformation due to rotation, where $\lambda = (R_p/R_e)^2$. It decreases as the star becomes more oblate: $f(1) = 1$, and $f(0) = 0.6$. 
The factor $k_n$ is a dimensionless structural constant (typically $\sim 1$) that depends on the star's density profile. For a polytropic equation of state, $P = K \rho^{1 + 1/N}$, it can be derived from the Lane-Emden function. For example, $k_n = 0.81482$ for $n = 0.5$, and $k_n = 0.65345$ for $n = 1$. If we assume that $q f(\lambda)/k_n \approx \rm const$ along the PNS's evolutionary stages, we can integrate Eq. \eqref{eqj1} to obtain
\begin{equation}
    J_f=J_i\left( \frac{M_{f}}{M_{i}}\right)^{q \frac{f(\lambda)}{k_n}} \label{eqj2}.
\end{equation}
We want to evaluate the situation were the loss of angular momentum by neutrino emission is maximized, {in this way}, we take the extreme case where $q=5/3$, $f(\lambda)=1$ and $k_n=0.65345$, and obtain the expression we are going to use in this work, given by 
\begin{equation}
    J_f=J_i\left( \frac{M_{f}}{M_{i}}\right)^{2.55} \label{eq1}
\end{equation}
where $J_{i}$ and $J_{f}$ are the initial and final angular momentum, respectively, and $M_{i}$ and $M_{f}$ are the initial and final gravitational mass, respectively. 

Additionally, we are going to investigate the energy distribution of the considered PNSs. Specifically, we are interested in the changes in the kinetic energy $T_{\rm kin}$, internal energy $U$ and gravitational binding energy $|W|$ energy as the PNSs evolve. For a stationary and uniformly rotating star, these quantities are given, respectively, by the following expressions~\cite{Stergioulas:2003yp}: 
\begin{equation}
    T_{\rm kin} = \frac{J \Omega}{2}, \qquad U = M_p - M_0 \quad \text{ and } \quad W = M - M_p - T,
\end{equation}
where, $M$ is the gravitational mass and $M_p$ is the proper mass and $M_0$ is the baryon mass.

\section{Evolution of intermediate mass rapidly rotating PNS} \label{sec_160}

\begin{table}[!ht]
 \caption{Rotational and compositional properties of evolving PNSs with baryonic mass $1.60\,M_\odot$.}
    \centering
    \setlength\extrarowheight{2pt}
    \begin{ruledtabular}
    \begin{tabular}{ c c c c c c c c c c c c c c }
   $s_B, \,Y_L$ & EoS &\multicolumn{4}{c}{$J$ conserved}&
\multicolumn{4}{c}{$J$ loss}&\multicolumn{4}{c}{static}\\
  \cline{3-6} \cline{7-10} \cline{11-14}
      &   & $J[M_\odot^2]$ & $n_c/n_0$ & $f_R[\%]$ & $M/R_e$ & $J[M_\odot^2]$ & $n_c/n_0$ & $f_R[\%]$ & $M/R_e$ & $J[M_\odot^2]$ & $n_c/n_0$ & $f_R[\%]$ & $M/R_e$ \\
        \hline
                & \ft N & 1.34 & 2.02 & 31 & 0.13 & 1.34 & 2.02  & 31 & 0.13 & 0 & 2.23 & 0  & 0.16  \\

        $1, 0.4$  & \ft NH & 1.34 & 2.02 & 30 & 0.13 & 1.34 & 2.02 & 30 & 0.13 & 0 & 2.24 & 0 & 0.16  \\

                & \ft NHD & 1.34 & 2.03 & 30 & 0.13 & 1.34 & 2.03 & 30 & 0.13 & 0 & 2.25 & 0 & 0.16   \\

        \hline
                & \ft N & 1.34 & 1.77 & 31 & 0.12 & 1.29 & 1.77 & 29 & 0.12 & 0 & 1.98 & 0 & 0.14  \\

        $2, 0.2$  & \ft NH & 1.34 & 1.85 & 31 & 0.12 & 1.31 & 1.86 & 30 & 0.12 & 0 & 2.10 & 0 & 0.15  \\

                & \ft NHD & 1.34 & 1.98 & 32 & 0.12 & 1.31 & 2.00 & 30 & 0.12 & 0 & 2.24 & 0 & 0.15   \\

        \hline 
                & \ft N & 1.34 & 1.77 & 32 & 0.11 & 1.25 & 1.78 & 30 & 0.12 & 0 & 1.98 & 0 & 0.14  \\

        $2, 0$  & \ft NH & 1.34 & 1.88 & 33 & 0.11 & 1.29 & 1.89 & 30 & 0.12 & 0 & 2.15 & 0 & 0.15  \\

                & \ft NHD & 1.34 & 2.12 & 51 & 0.10 & 1.29 & 2.15 & 32 & 0.12 & 0 & 2.37 & 0 & 0.15   \\

        \hline
                & \ft N & 1.34 & 2.09 & 30 & 0.14 & 1.25 & 2.10 & 28 & 0.14 & 0 & 2.30 & 0 & 0.16  \\

        $T=0$  & \ft NH & 1.34 & 2.09 & 30 & 0.14 & 1.29 & 2.10 & 28 & 0.14 & 0 & 2.30 & 0 & 0.16  \\

                & \ft NHD & 1.34 & 2.17 & 30 & 0.14 & 1.29 & 2.20 & 27 & 0.14 & 0 & 2.42 & 0 & 0.16  \\          
    \end{tabular}
    \end{ruledtabular}
    \label{tab1}
\end{table}
In this section, we study the properties of a rapidly rotating PNS that has a baryonic mass of $M_0=1.60$~\msun, which is conserved along the star's evolution. We consider three possible evolutionary scenarios: one in which the angular momentum $(J)$ of the PNS is conserved, represented by solid lines in our plots, the case where the star loses angular momentum due to neutrino emission, represented by dashed lines in our plots, and also a scenario where the star is static, represented by dash-double-dot lines in our plots. In the first part of this work~\cite{daSilva:2025cfe}, we concluded that for a PNS with $M \approx 1.5$~\msun~the EoS that leads to the smallest values of Keplerian frequency $(\Omega_K)$ is the EoS with nucleons plus hyperons and $\Delta$-resonances (NHD). In addition, we determined that the stage where the star becomes neutrino transparent and reaches its highest temperature $(s_B=2, \,Y_L=0)$ is the stage when $\Omega_K$ is smaller. For this reason, we chose the angular momentum of our PNS to be the one that leads the star to rotate with $\Omega_K$ when we consider the EoS NHD at the stage $s_B=2, \,Y_L=0$. Otherwise, the star would exceed the mass-shedding limit at this critical evolutionary phase. With this choice, we can study a PNS rotating near $\Omega_K$ in a physically viable rotational configuration across all evolutionary stages and EoSs considered, avoiding scenarios where the star would become unstable due to excessive angular momentum. This way, for the two scenarios where the PNS is rotating, we assume that the star has an angular momentum of $J=1.34\, M_{\odot}^2$ in the first snapshot $(s_B=1, \,Y_L=0.4)$. In the scenario of conserved angular momentum, the value of $J$ remains constant in all evolutionary stages, while for the scenario with loss of angular momentum, the value of $J$ decreases according to Eq. \eqref{eq1}.

\Cref{tab1} summarizes the results for some of the rotational properties of the PNS considered in this section along its evolutionary stages, using the static case as a baseline. The table shows that deformation increases when the star's angular momentum $J$ is conserved, compared to the case where the star loses angular momentum through neutrino emission. This effect becomes more pronounced during the deleptonization and neutrino-transparent stages, when the star is losing neutrino pressure support. Additionally, this behavior is accompanied by a slight increase in the central baryon density $n_c$ in the angular momentum loss scenario, which increases significantly when we consider the star in the static configuration, with $J = 0$. Generally, we found that the star experiences its highest deformation rate during the intermediate stages of stellar evolution, when the stellar matter is hot, particularly in the neutrino-transparent stage when the matter reaches its peak temperature. Thus, deformation is linked to an increase in thermal pressure. In contrast, we found less deformation in all cases analyzed when the star is cold and catalyzed. Additionally, an increase in the degrees of freedom of the stellar matter was observed to drive deformation, as seen in the intermediate stages, with the deformation reaching $51\%$ for NHD-admixed hypernuclear matter in the neutrino-transparent phase. This pronounced deformation is also associated with the PNS reaching the Keplerian frequency for this specific configuration, which is associated with the maximum deformation a star can have before it starts shedding its mass. Comparing the three EoSs under the condition of angular momentum loss, we find that the NH and NHD compositions experience a total angular momentum reduction of $3.7\%$ along the evolution of the PNS, whereas the NHD EoS shows a $6.7\%$ decrease.
\begin{figure}[t!]
    \centering
    \includegraphics[width=0.49\linewidth]{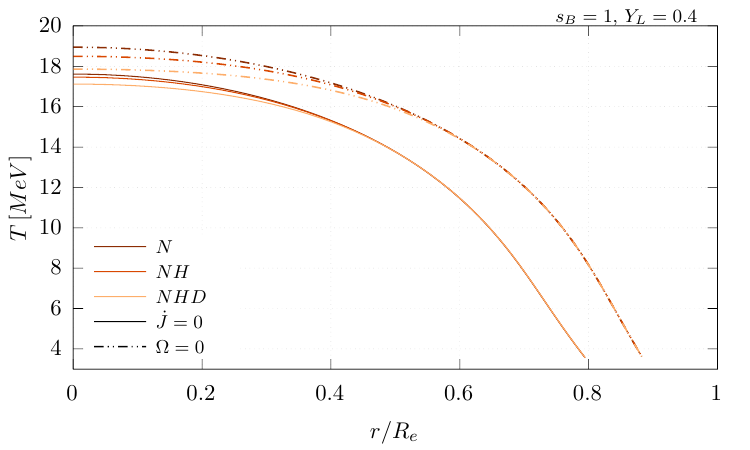}
    \includegraphics[width=0.49\linewidth]{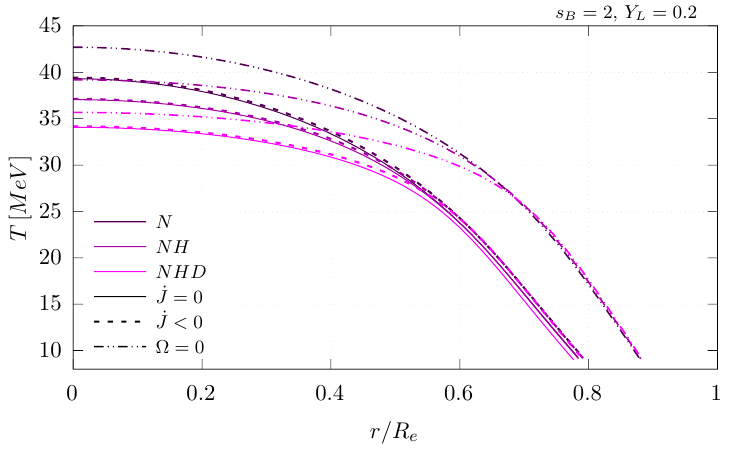}
    \includegraphics[width=0.49\linewidth]{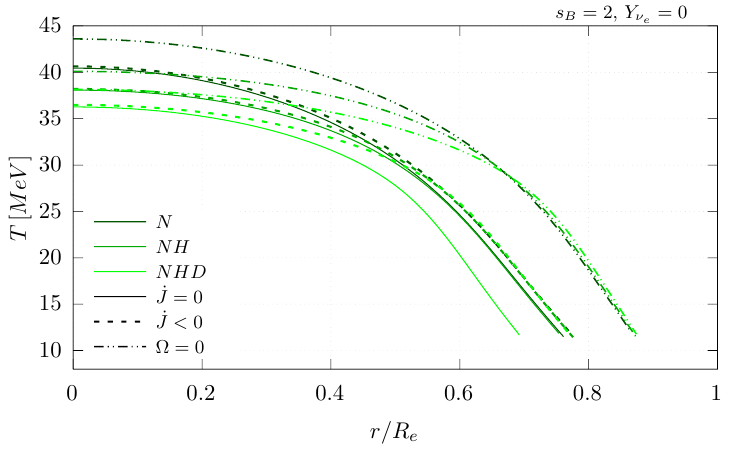}
    \caption{Profiles of temperature versus normalized {equatorial plane radius, $r/R_e$,} for a star with $M_0 = 1.60$~\msun, solid lines are for conserved angular momentum, dashed lines are for loss of angular momentum by neutrino emission, and dash-double-dot lines are for static stars. Top left: $s_B=1$ and $Y_L=0.4$, top right:  $s_B=2$ and $Y_L=0.2$, bottom: $s_B=2$ and $Y_{\nu_e}=0$.}
    \label{figrext}
\end{figure}

\Cref{figrext} shows temperature profiles plotted against the normalized equatorial radius for a $M_0=1.60$~\msun~PNS evolving from birth to the neutrino-transparent stage \cite{Pons:1998mm}. Three matter compositions are compared: nucleonic \cite{Lattimer:2000nx}, hyperon-admixed \cite{Weissenborn:2011kb}, and matter including additional $\Delta$-resonances \cite{Issifu:2023qyi}. Rotational states are distinguished by line style: solid lines represent configurations with conserved angular momentum ($J = \text{const}$) \cite{Stergioulas:2003yp}, dashed lines correspond to models with angular momentum loss ($\dot{J} < 0$), and dash-double-dot lines indicate static stars ($\Omega = 0$). The results show that $J = \text{const}$ models maintain systematically lower core temperatures due to centrifugal support, which reduces the $n_c$. In contrast, static configurations reach higher $n_c$ (see \Cref{tab1}) and thus higher temperatures~\cite{Glendenning2000}. This thermal contrast arises from rotational flattening, which increases the equatorial radius and redistributes thermal energy across a larger volume, thereby reducing the temperature gradient, with $|\nabla T| \propto R_e^{-1}$ \cite{Camelio:2019rsz}. Across all panels, stars composed purely of nucleons exhibit the highest temperature profiles. Introducing hyperons reduces the temperature \cite{Oertel:2016bki}, and the inclusion of $\Delta$-resonances lowers it further \cite{Issifu:2023qyi}. This trend reflects the role of additional degrees of freedom, which distribute the available thermal energy among more baryonic species, effectively reducing the thermal energy per baryon.

Comparing the first panel (initial stage) with the second and third panels (intermediate stages) reveals that differences in the temperature profiles among the three cases (conserved angular momentum, momentum lost, and static) become increasingly pronounced as the star evolves. This is primarily due to the stabilizing effect of neutrino pressure during the early, lepton-rich phase, which helps suppress rotational instabilities and limits thermal variations \cite{1986ApJ...307..178B}. As deleptonization progresses, this stabilizing influence diminishes, allowing rotation to exert a greater impact on both the thermal and structural evolution of the star \cite{Stergioulas:2003yp}. Moreover, examining the magnitude of the temperature profiles across the panels shows that in the first stage, characterized by a high $Y_L$ and low $s_B$, the temperatures remain relatively low \cite{Roberts:2016mwj}. In contrast, during the second (deleptonization) and third (neutrino-transparent) stages, as $Y_L$ decreases and eventually vanishes $Y_{\nu_e} = 0$, the star attains higher entropy levels, resulting in elevated temperature profiles.
\begin{figure}[t!]
    \centering
    \includegraphics[width=0.49\linewidth]{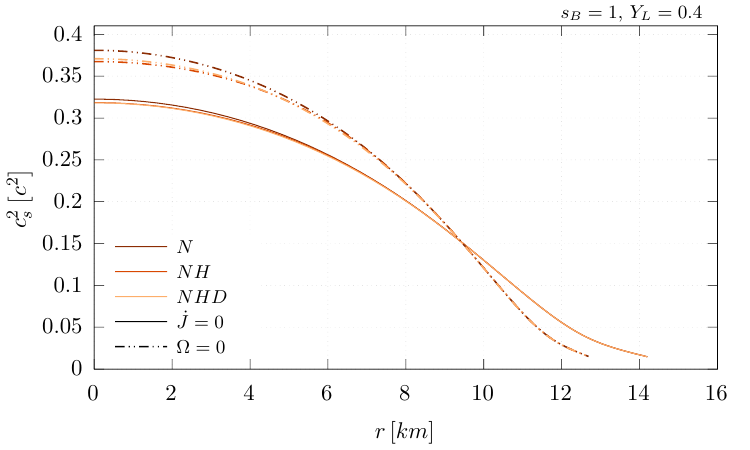}
    \includegraphics[width=0.49\linewidth]{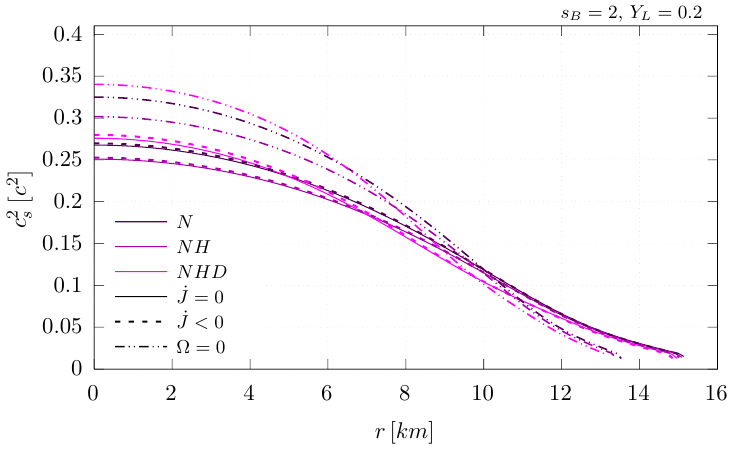}
    \includegraphics[width=0.49\linewidth]{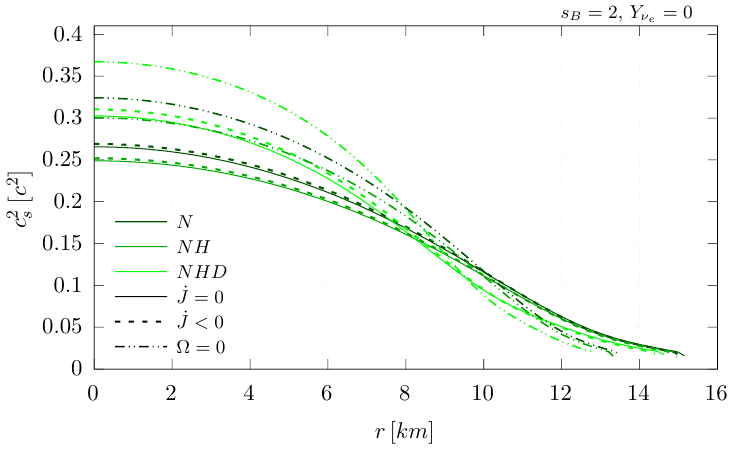}
    \includegraphics[width=0.49\linewidth]{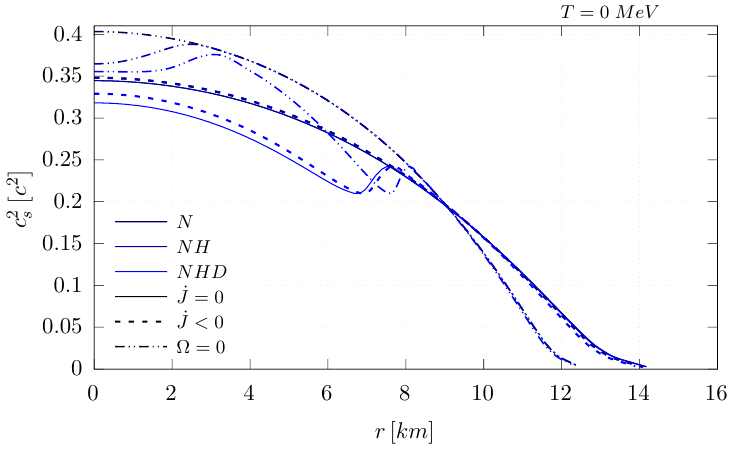}
    \caption{Profiles of squared sound velocity versus the equatorial plane radius, $r$, for a star with $M_0 = 1.60$~\msun, solid lines are for conserved angular momentum, dashed lines are for loss of angular momentum by neutrino emission, and dash-double-dot lines are for static stars.}
    \label{figrexcs}
\end{figure}

In \Cref{figrexcs}, we show the profiles of the squared sound speed, $c_s^2$, for an evolving PNS with a baryon mass $M_0 = 1.60$~\msun. The magnitude of $c_s^2$ serves as a sensitive probe of the stiffness of the underlying EoS \cite{Bedaque:2014sqa}: stiffer EoS produces higher $c_s^2$ values, particularly toward the stellar center, while softer EoS yields lower values \cite{Braun:2022jme, Issifu_20251}. Moreover, we observe that stars composed purely of nucleons exhibit the highest $c_s^2$. As hyperons are introduced, the magnitude decreases, with a further reduction observed when $\Delta$-resonances are included. This trend reflects the progressive softening of the EoS as additional degrees of freedom are incorporated into the stellar matter \cite{Issifu:2024fuw}.

Across the panels, we observe that $c_s^2$ is lowest in the case of conserved angular momentum (solid lines), increases moderately when angular momentum is lost via neutrino emission (dashed lines), and reaches its maximum in the static configuration (dash-double-dot lines). This behavior arises because centrifugal flattening in rotating stars lowers the central density, $n_c$, shifting the peak of $c_s^2$ outward to larger radii compared to static stars \cite{Friedman_Stergioulas_2013, Paschalidis:2016vmz, 1994ApJ...424..823C}. This highlights how rotation effectively softens the EoS by redistributing matter away from the core \cite{Haensel:2007yy}. As the star begins to lose angular momentum, the $c_s^2$ profile gradually transitions toward that of the static configuration. The degree of this shift depends on the interplay between neutrino viscosity and the timescales of angular momentum transport \cite{1986ApJ...307..178B, Thompson:2004if}. The higher $c_s^2$ observed in the static case reflects the greater central compactness and serves as a reference baseline for quantifying rotational effects. Overall, rotation suppresses $c_s^2$ in the stellar core due to centrifugal forces, an effect that becomes more pronounced at higher rotation rates \cite{Friedman_Stergioulas_2013, Lattimer:2000nx}.

\begin{figure}[t!]
    \centering
    \includegraphics[width=0.49\linewidth]{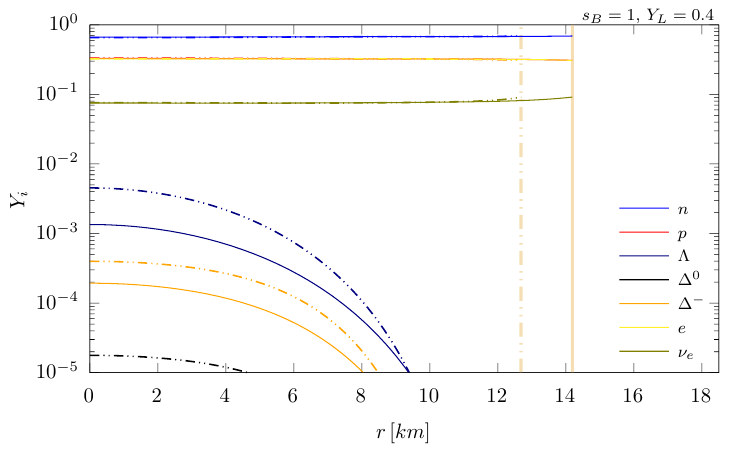}
    \includegraphics[width=0.49\linewidth]{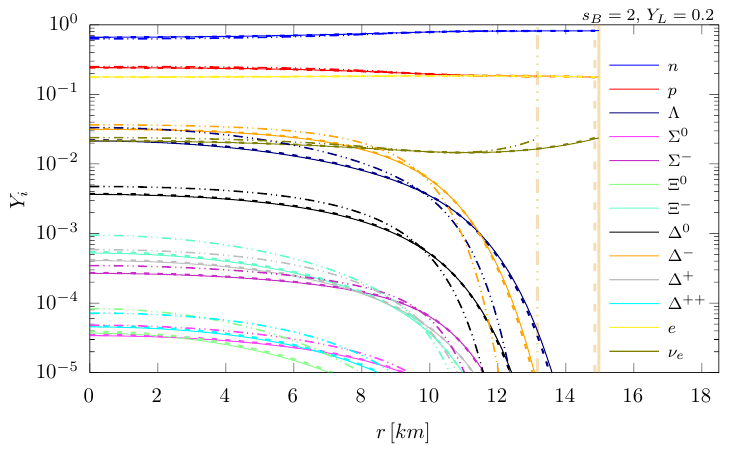}
    \includegraphics[width=0.49\linewidth]{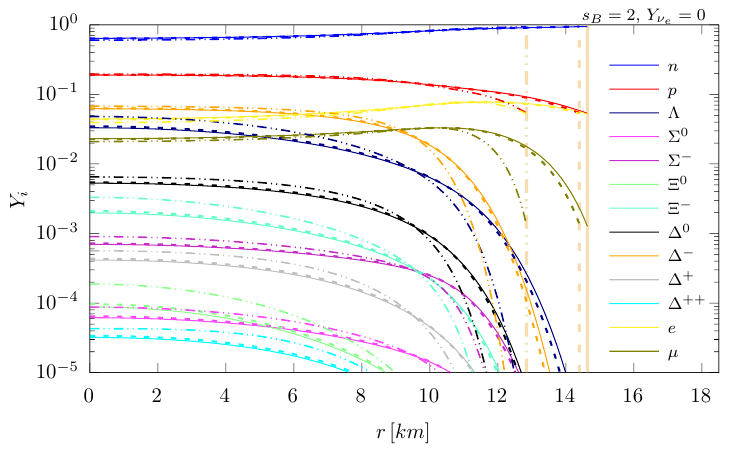}
    \includegraphics[width=0.49\linewidth]{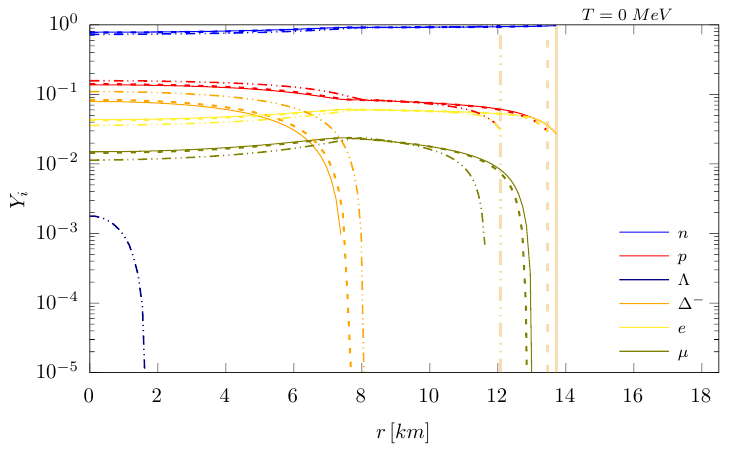}
    \caption{Profiles of particle distribution versus the equatorial {plane} radius for a star with $M_0 = 1.60$~\msun~and considering the EoS with nucleons plus hyperons and delta resonances (NHD). Solid lines are for conserved angular momentum, dashed lines are for loss of angular momentum by neutrino emission, and dash-double-dot lines are for static stars. Vertical lines represent the radius where the stellar crust begins. }
    \label{figrexyi}
\end{figure}

Across the panels, we observe that the impact of rotation on $c_s^2$ is smaller during the initial stage of evolution, largely because neutrino pressure provides significant support against rotational instabilities \cite{Pons:1998mm}. As the star deleptonizes (second panel, top) and eventually becomes neutrino-transparent (first panel, bottom), the influence of rotation grows more pronounced \cite{Glendenning2000}. This results in a distinct separation between profiles with conserved angular momentum, characterized by centrifugal flattening and reduced central densities, and those experiencing angular momentum loss, which evolve toward more compact, static-like configurations. In the cold, catalyzed stage (second panel, bottom), we note prominent bumps in the $c_s^2$ profiles for hyperon-admixed matter and for matter further enriched with $\Delta$-resonances, reflecting the emergence of stratified particle layers in the stellar core.  These features arise because the appearance of new particle species alters the local stiffness of the EoS, creating discontinuities or sharp changes in the sound speed. The radial positions of these bumps precisely match the density thresholds at which specific particles emerge, as shown in \Cref{figrexyi}, confirming the direct link between the microphysical composition and variations in the sound-speed profile (see references \cite{Issifu_20251, Issifu:2024fuw} and references therein for more details). In the fixed entropy state, these discontinuities are smoothed out because finite temperature effects and thermal fluctuations broaden particle population thresholds and create a more homogeneous matter distribution throughout the stellar interior. 

In \Cref{figrexyi} we show the particle distributions in the stellar matter as a function of the equatorial radius for the PNS considered in this section. In this figure, we consider the EoS composed of nucleons plus hyperons plus delta isobars (NHD). We observed that in the first snapshot of evolution (top left panel) of a PNS with $M_0 = 1.60$~\msun~the only hyperon that appears in the matter is $\Lambda$, for the rotating case only the $\Delta^-$ appears, and in the static configuration we observed a small amount of $\Delta^0$ baryons appearing towards the core of the star. For the rotating case, the fraction of $\Lambda$ and $\Delta^0$ is smaller than in the static case, which is consistent with the rotation effect increasing the stellar radius and decreasing the core density, which is required to produce heavy baryons. However, at this stage, we cannot observe any significant difference in the amount of nucleons, electrons, and neutrinos when we compare static and rotating cases. Besides, the fraction of $e$ and $p$ is practically identical throughout the star.

In the second stage of evolution (top right panel) of \Cref{figrexyi}, when the lepton fraction has decreased to $Y_L=0.2$ and the temperature of the star has significantly increased, as shown in \Cref{figrext}. As expected, the amount of neutrinos and electrons is considerably smaller at this stage than in the previous one, and in both the static and rotating cases, the amount of neutrinos slightly increases near the surface of the star. We also observe that due to the heating of the star, all deltas and almost all hyperons considered in our EoS are present in the stellar matter in substantial quantities at this stage. Additionally, all heavy baryons present tend to appear closer to the crust of the star than in the first stage of evolution. Additionally, all the hyperons and the deltas present are more abundant in the static case than in the rotating ones, especially towards the core of the star. At this stage, we cannot observe any significant difference between the scenario with conserved angular momentum and the one with loss of angular momentum due to neutrino emission. 

Looking at the third snapshot (bottom left) of \Cref{figrexyi}, when the star becomes neutrino transparent $(Y_{\nu_e}=0)$ and the temperature in the stellar matter reaches its peak. At this stage, we observe a decrease in the electron abundance and, in the absence of neutrinos, the creation and accumulation of muons becomes energetically favorable to maintain charge neutrality and beta equilibrium~\cite{Sugiura2022PTEP, Issifu:2024fuw}.
We also observe that the rise in temperature leads to the proportion of all hyperons and of $\Delta^0$ and $\Delta^-$ increasing even further than in the previous snapshot, and they also start to appear closer to the crust than before. This points to thermal production of particles. On the other hand, in the static case, the proportion of the two positively charged delta baryons ($\Delta^+$ and $\Delta^{++}$) decreases, whereas in the rotating case, only the proportion of $\Delta^{++}$ decreases compared to the previous stage of evolution. This difference can be attributed to the condition of charge neutrality inside the star, since in the static configuration, the number of electrons and muons near the core is lower than in the rotating case. We also observe that, even at this stage, the $\Sigma^+$ baryons do not appear in significant amounts inside a star with $M_0 = 1.60$~\msun. Again, the proportion of hyperons and deltas is higher in the static case than in the rotating ones, especially towards the center of the star. Additionally, at this stage, no significant difference in the particle fractions between the case with and without angular momentum conservation is observed.

In the bottom right panel of \Cref{figrexyi}, we show the last stage of evolution when the star becomes a cold and catalyzed NS. In this panel, we observe that a NS described by the NHD EoS and with $M_0 = 1.60$~\msun~does not contain any hyperons in significant amounts if it is rotating close to the Keplerian limit. In contrast, if the NS is static, it contains only $\Lambda$ hyperons. In both cases, rotating and non-rotating, only the $\Delta^-$ resonances appear in the stellar matter. At this stage, rotation slightly increases the fractions of $\mu$, $e$, and $n$, which are the particles favored by the decreasing central density, while the fractions of $p$, $\Delta^-$, and $\Lambda$ decrease. Finally, we notice a small difference in the fraction of delta baryons when comparing the cases with and without conservation of angular momentum. However, we can conclude that for a PNS with $M_0 = 1.60$~\msun~the case with loss of angular momentum due to neutrino emission does not lead to any significant difference in the analysis of particle fractions.
\begin{figure}[t!]
    \centering
    \includegraphics[width=0.49\linewidth]{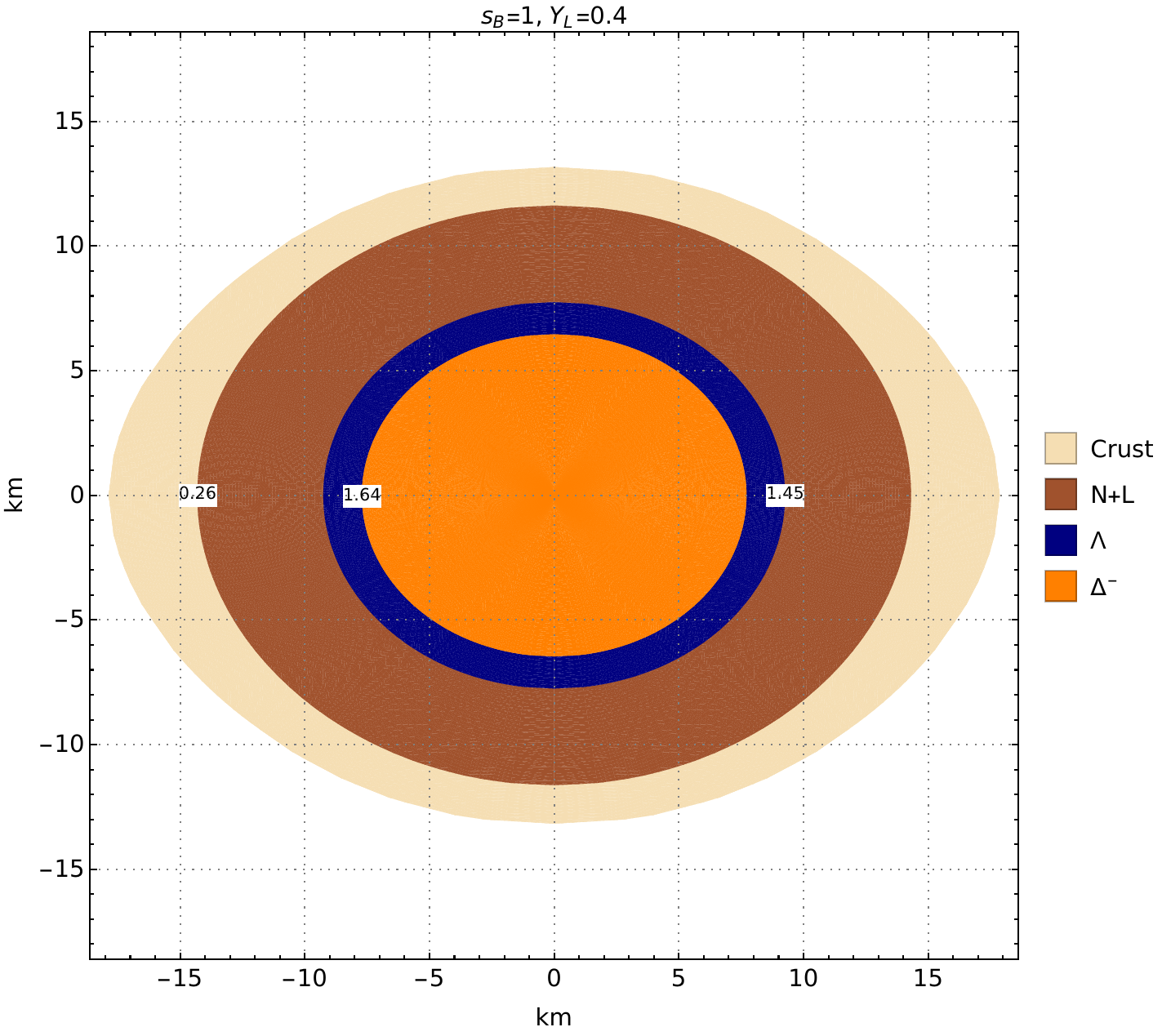}
    \includegraphics[width=0.49\linewidth]{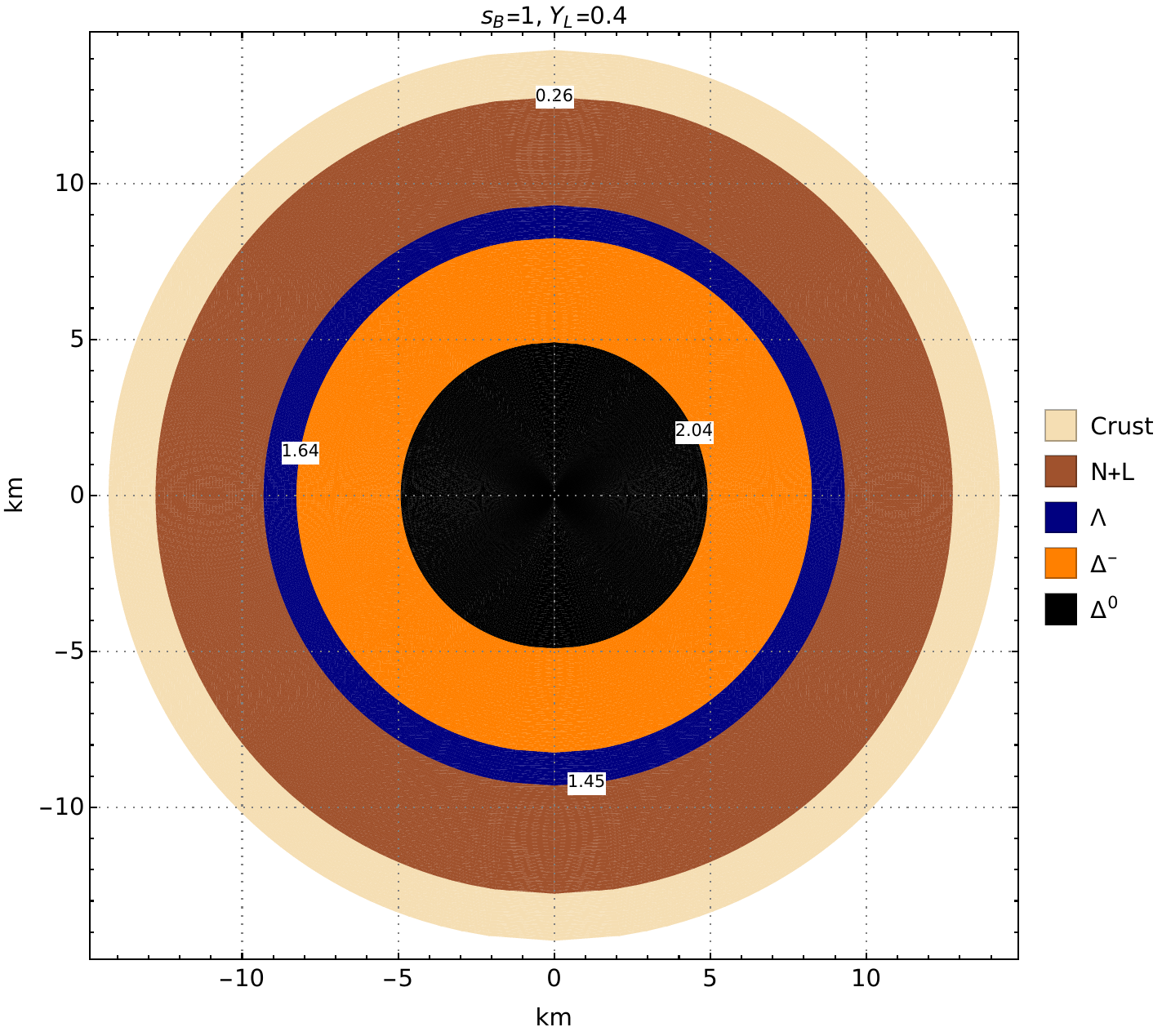}
    \includegraphics[width=0.49\linewidth]{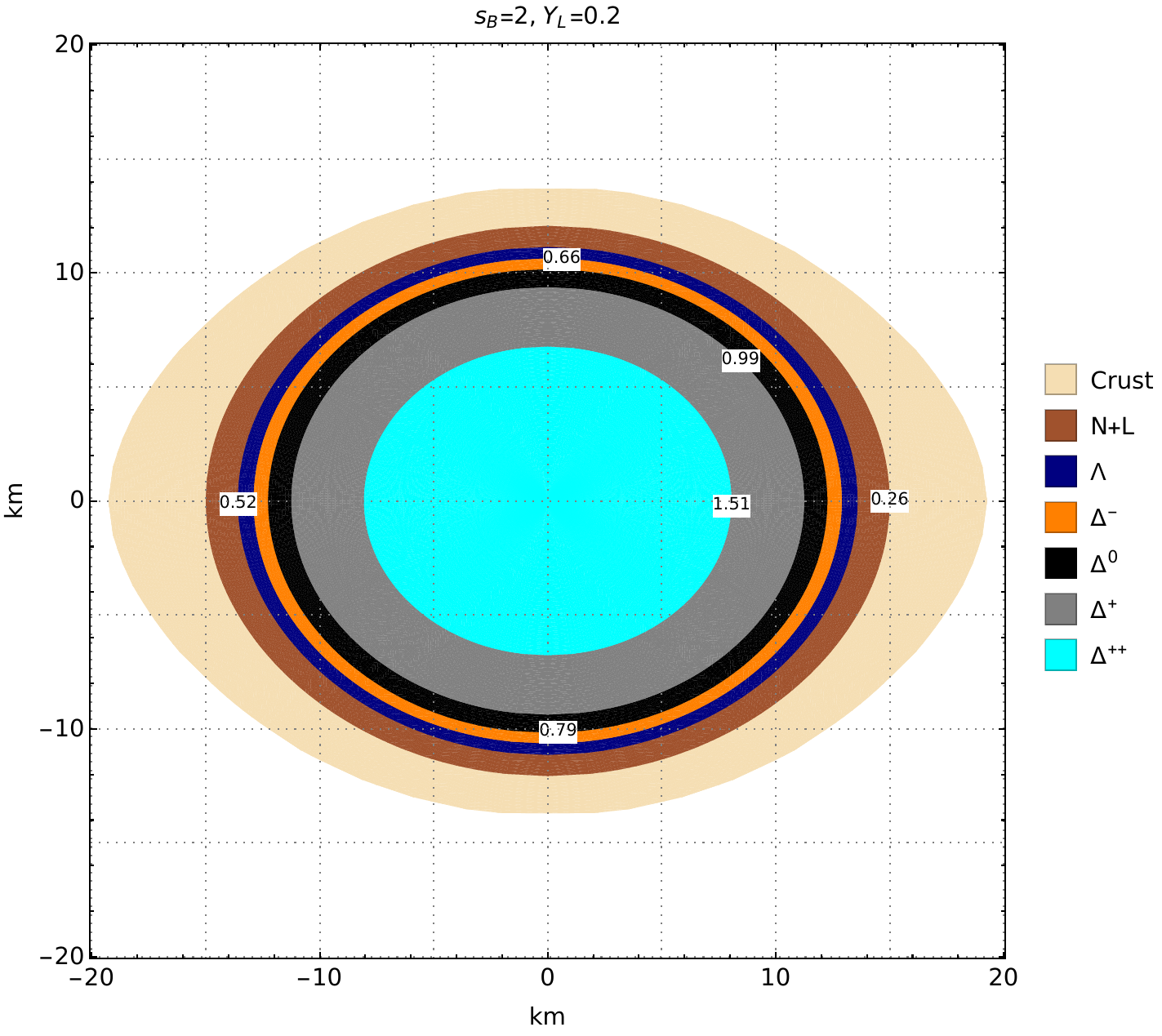}
    \includegraphics[width=0.49\linewidth]{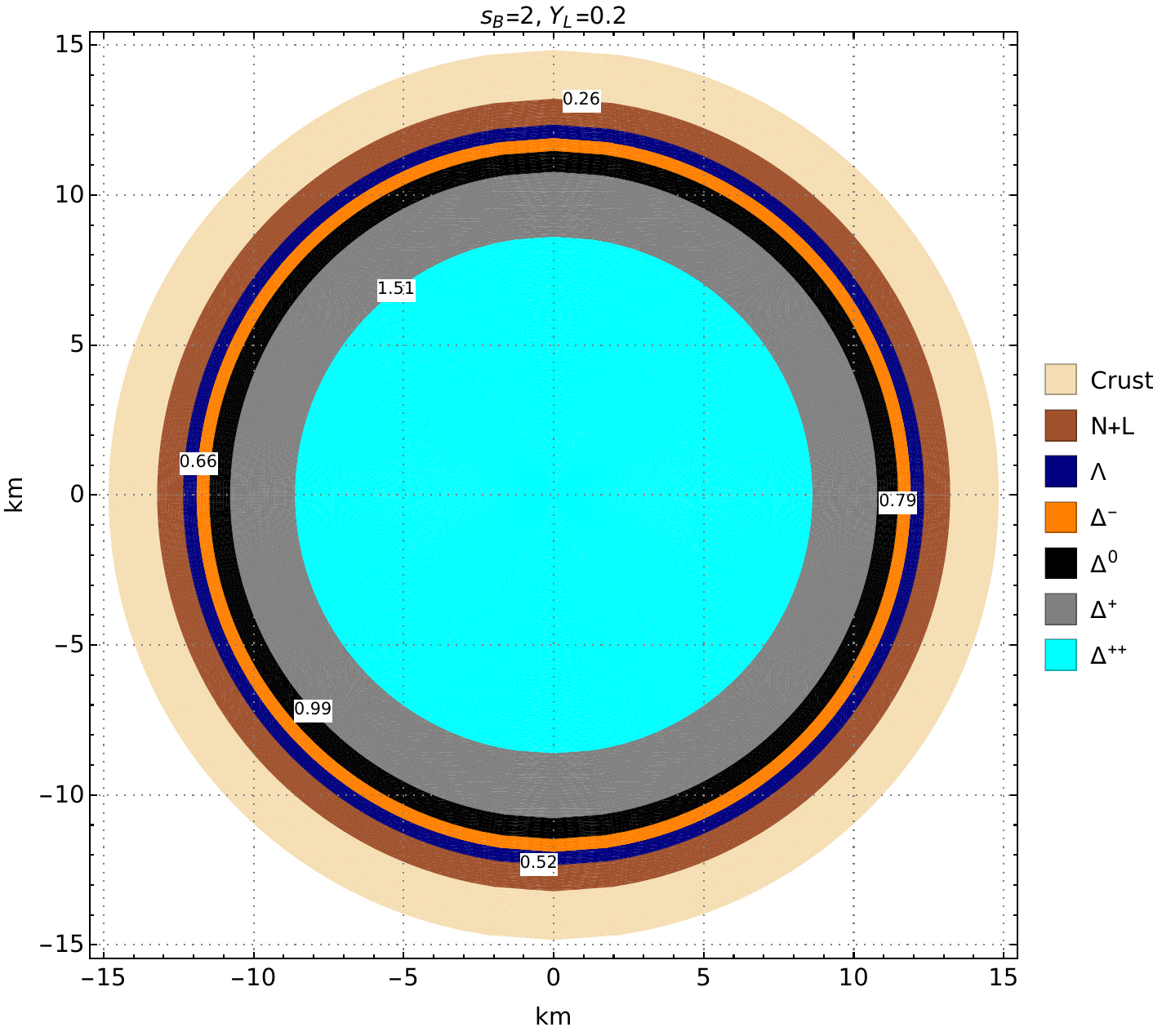}
    \caption{Contour plots of a PNS with a baryonic mass of $M_0 = 1.60$~\msun, composed of nucleons, hyperons and delta resonances (NHD EoS), during the neutrino opaque stages of evolution. The evolutionary stage of each panel is indicated at the top, while the legend for the meaning of each color is shown on the right. The left panels show the rapidly rotating scenario with conserved angular momentum, while the right panels represent the static scenario. From the surface toward the center of the star, each layer corresponds to the radial position where a specific type of particle begins to appear in the stellar matter. ``N+L'' denotes nucleons plus leptons. The values indicated along each contour represent the baryon density $(n_B/n_0)$, at that location.}
    \label{fignB1}
\end{figure}

\Cref{fignB1} shows contour plots for different stages of PNS evolution, with the $\dot{J} = 0$ case presented in the left panel and the $\Omega = 0$ case in the right panel, covering both the neutrino-trapping and deleptonization stages. These figures indicate the radius and baryon density at which $\Delta$ baryons, $\Lambda$ hyperons, and the stellar crust first appear. The full particle composition of the PNS is shown in \Cref{figrexyi}. In the first stage of evolution (top panels), the crust of the star appears at $n_B = 0.26\,n_0$. In the static configuration, the crust is symmetric, whereas in the rapidly rotating case, it becomes deformed, thinner at the poles and thicker at the equator. In the rotating case, only $\Delta^-$ baryons are present at the core, while in the static case, where the central density is higher, $\Delta^0$ baryons also appear at the core.
In the second evolutionary snapshot (bottom panels), heating of the PNS causes the stellar radius to increase and the central baryon density to decrease in both scenarios, compared to the previous stage. The reduction in $n_c$ is more pronounced in the rotating case because, in addition to heating, the rotation rate increases at this stage. Despite this, all $\Delta$ baryons considered in the EoS appear inside the star during this stage. This occurs because all particles begin to appear at lower densities in the second snapshot, for instance, $\Delta^-$ baryons, which first appear at $n_B = 1.64\, n_0$ in the initial stage, already appear at $n_B = 0.66\, n_0$ in the second stage.
\begin{figure}[t!]
    \centering
    \includegraphics[width=0.49\linewidth]{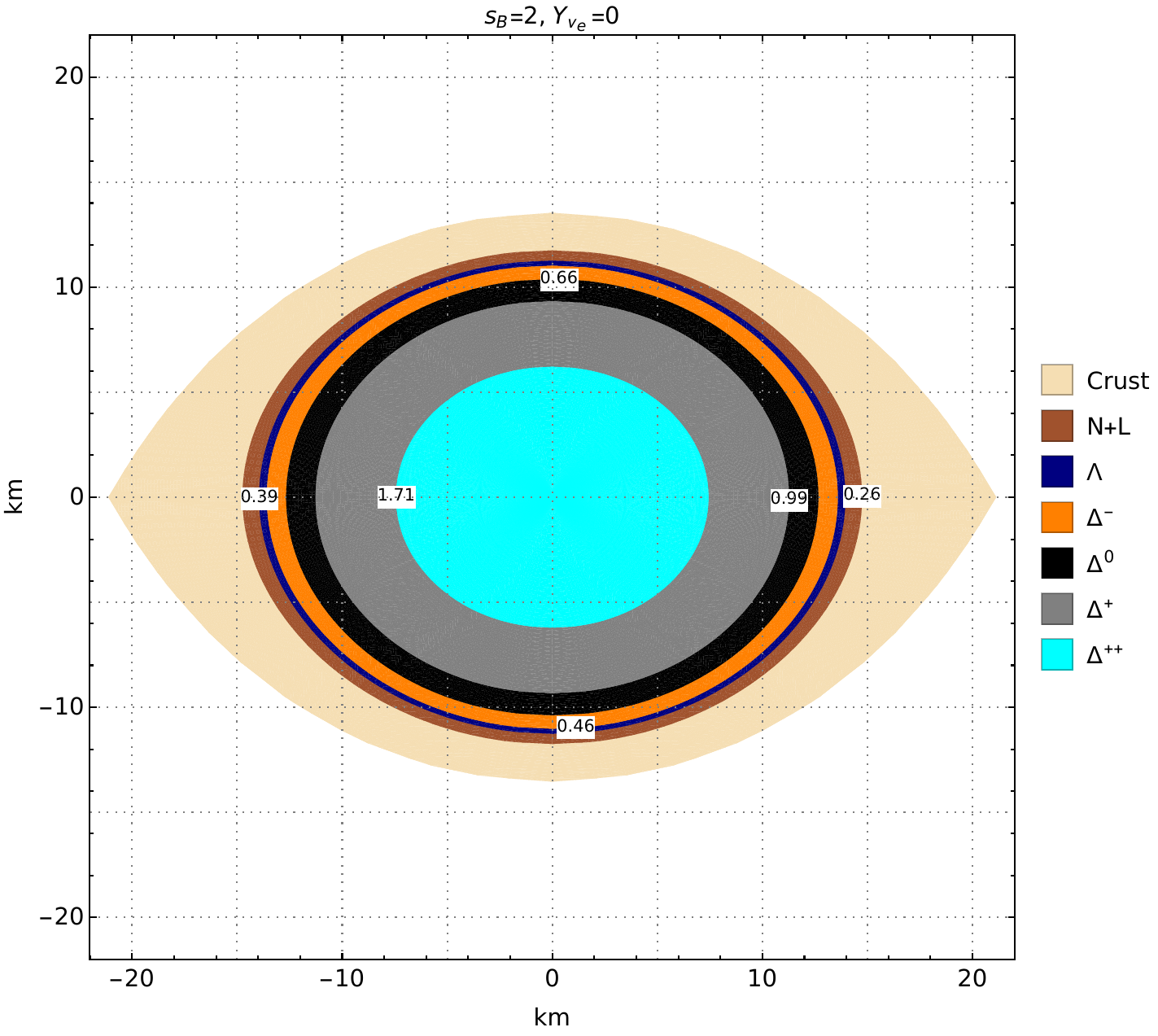}
    \includegraphics[width=0.49\linewidth]{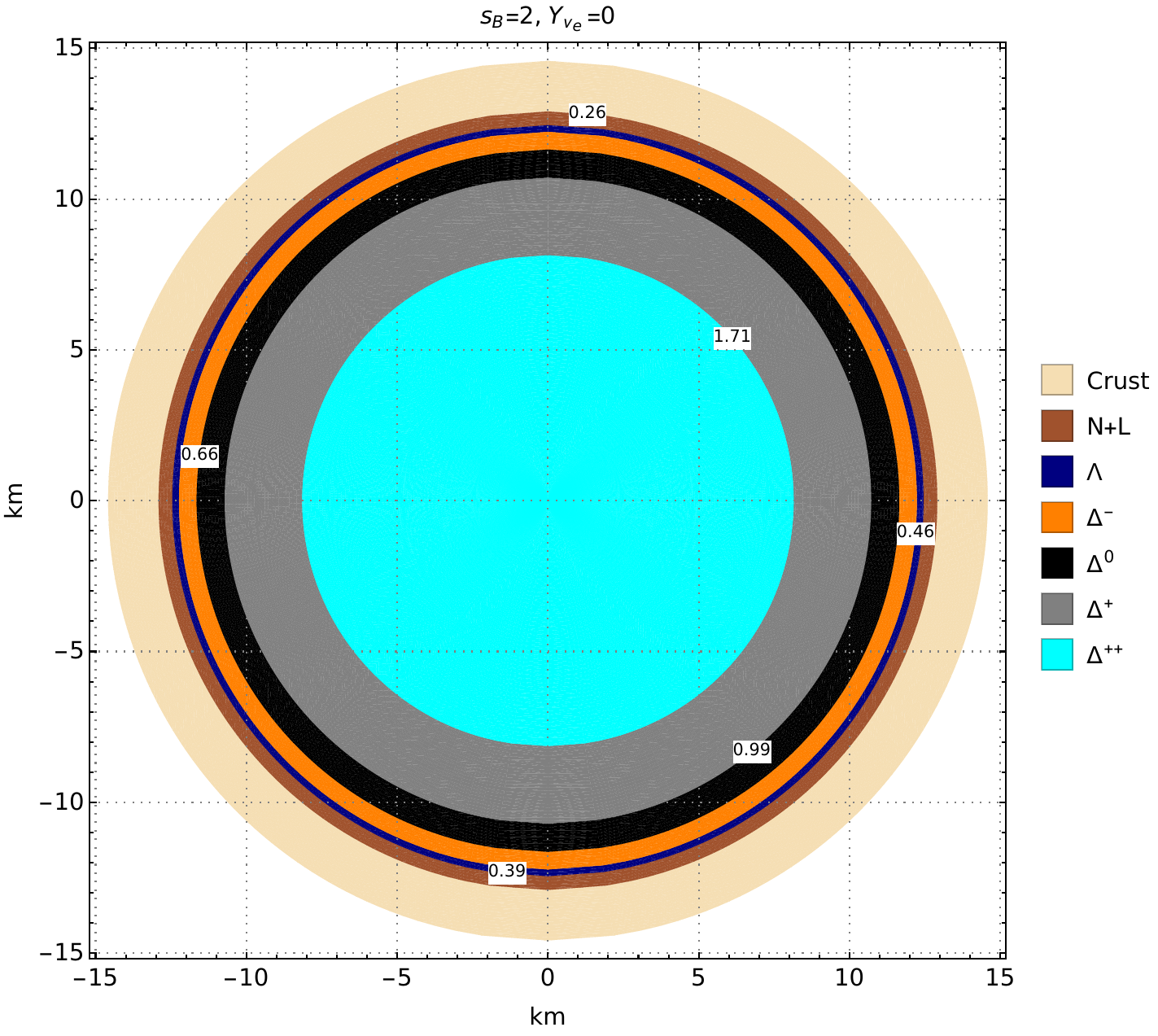}
    \includegraphics[width=0.49\linewidth]{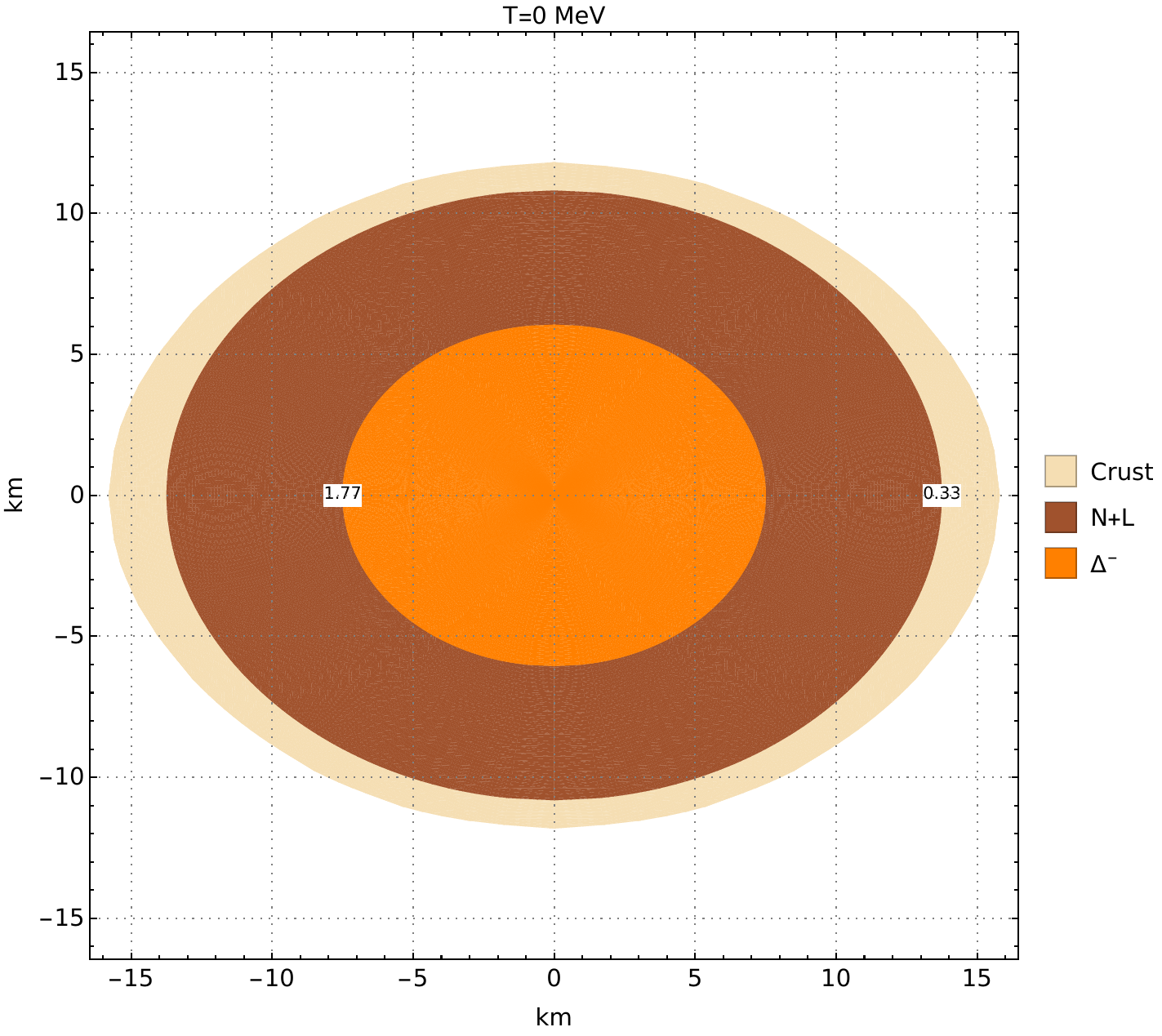}
    \includegraphics[width=0.49\linewidth]{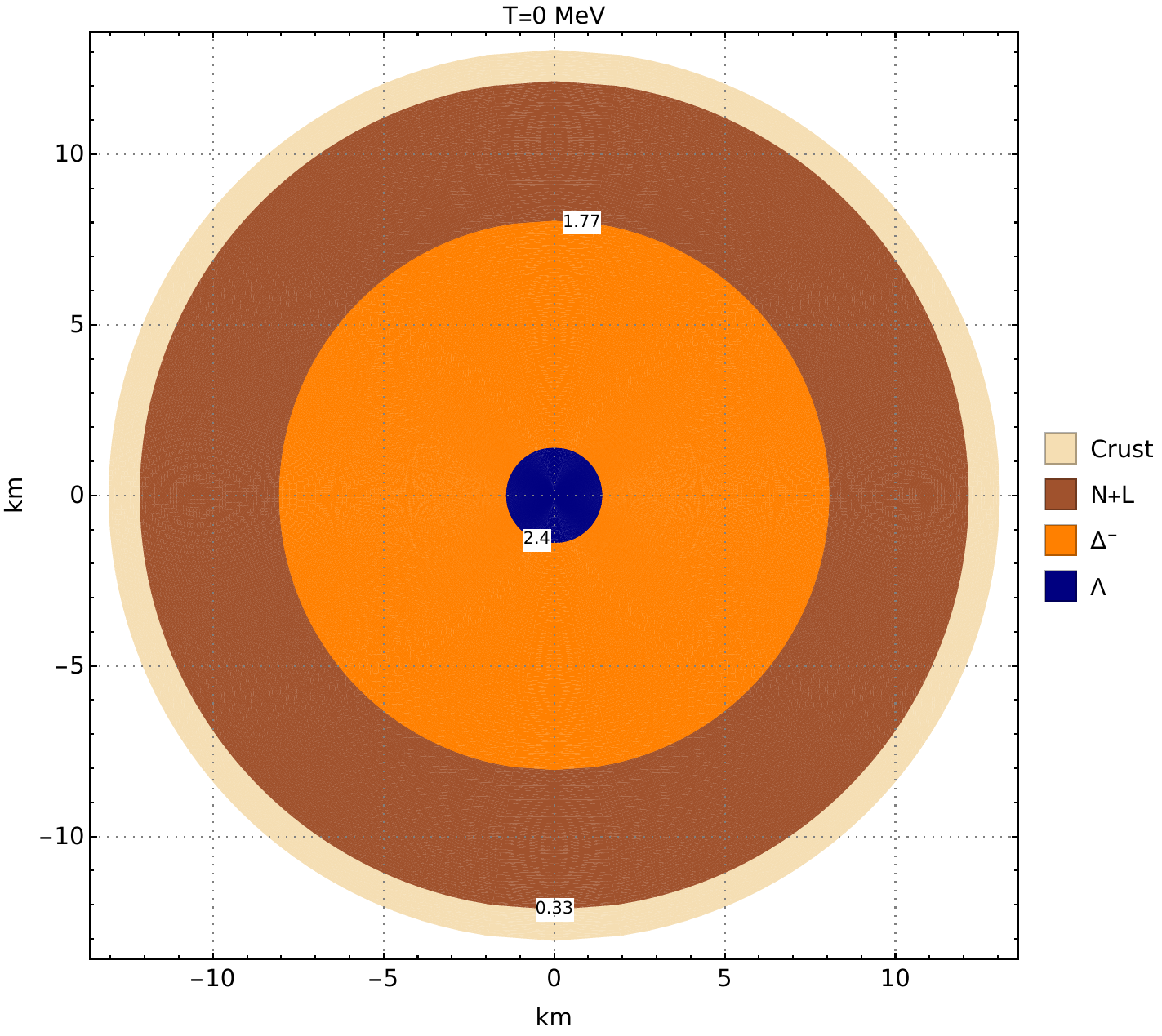}
    \caption{Contour plots of a PNS with a baryonic mass of $M_0 = 1.60$~\msun, composed of nucleons, hyperons and delta resonances (NHD EoS), during the neutrino transparent stages of evolution. The evolutionary stage of each panel is indicated at the top, while the legend for the meaning of each color is shown on the right. The left panels show the rapidly rotating scenario with conserved angular momentum, while the right panels represent the static scenario. From the surface toward the center of the star, each layer corresponds to the radial position where a specific type of particle begins to appear in the stellar matter. ``N+L'' denotes nucleons plus leptons. The values indicated along each contour represent the baryon density $(n_B/n_0)$, at that location.}
    \label{fignB2}
\end{figure}

\Cref{fignB2} shows contour plots of the PNS considered in this section, with the conserved angular momentum scenario ($\dot{J}=0$) presented in the left panel and the static scenario ($\Omega=0$) in the right panel, during the neutrino-transparent stages. As in \Cref{fignB1}, we indicate the equatorial radius and baryon density at which $\Delta$ baryons, $\Lambda$ hyperons, and the stellar crust first appear. At the stage when the PNS becomes neutrino transparent (top panels), the $\dot{J}=0$ configuration reaches the Keplerian frequency $\Omega_K$, thereby achieving its maximum possible equatorial radius. In contrast, in the $\Omega=0$ scenario, the radius of the PNS decreases slightly. In this snapshot, some particles appear at even lower densities than in the previous stage, for example, $\Delta^-$ baryons are now observed at $n_B = 0.46\,n_0$, while others, such as $\Delta^{++}$, appear at higher densities than before. In the bottom panels of \Cref{fignB2}, we show the final stage of the PNS’s life, when it has already cooled into a NS. At this stage, the stellar crust forms at a higher density ($n_B = 0.33\,n_0$), and is thinner than in earlier stages. In the rotating case, the star is less deformed than during the $s_B = 2$ stages, despite having a higher rotational rate. As already noted in \Cref{figrexyi}, in both scenarios the only $\Delta$ resonances present in significant amounts inside this star with $M_0 = 1.60$~\msun~are the $\Delta^-$ baryons, and only in the static case do $\Lambda$ hyperons appear, and then only very close to the stellar core.
\begin{figure}[t!]
    \centering
    \includegraphics[width=0.49\linewidth]{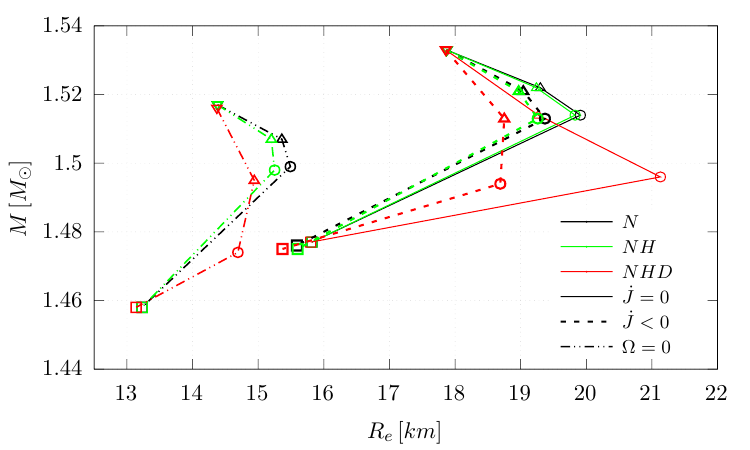}
    \includegraphics[width=0.49\linewidth]{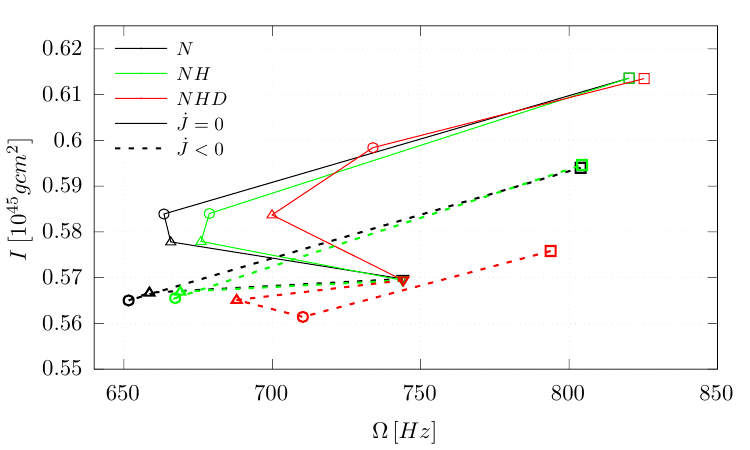}
    \caption{Evolution of the gravitational mass and equatorial radius, on the left, and of the moment of inertia and angular frequency, on the right, for a NS with $M_0 = 1.60$~\msun. Solid lines are for conserved angular momentum, dashed lines are for loss of angular momentum by neutrino emission, and dash-double-dot lines are for static stars. Inverted triangles are for $s_B=1, \,Y_L=0.4$, triangles are for $s_B=2, \,Y_L=0.2$, circles are for $s_B=2, \,Y_{\nu_e}=0$ and squares are for $T=0$ MeV.}
    \label{figpvst}
\end{figure}

On the left panel of \Cref{figpvst}, we track the changes in the gravitational mass and equatorial radius of a PNS with baryonic mass of $M_0=1.60$~\msun~in the three scenarios studied in this work: non-rotating (dash-double-dot lines) and rotating with (solid lines) and without (dashed lines) angular momentum conservation. In general, we observe that in all scenarios, the gravitational mass decreases during the evolution of a PNS. This reduction occurs because, as the star evolves, it undergoes deleptonization and loses substantial energy through neutrino emission. Without further accretion, these processes result in a less massive and more compact NS \cite{burrows1986, Pons:1998mm}. At the same time, the equatorial radius increases as the star deleptonizes and heats up, and then decreases again as the star becomes a cold and catalyzed NS. In general, the gravitational mass in the static scenario is a little smaller in size than in the rotating scenarios, since part of the gravitational mass is due to the rotational energy. 

On the other hand, the differences in radius between rotating and non-rotating cases are very pronounced because in the rotating scenarios we are considering stars rotating near the mass shedding limit. It is interesting to notice that, in each of the three scenarios, for a PNS with $M_0 = 1.60$~\msun~the EoS NH does not lead to any significant difference in $M$ or $R_e$ when compared to the EoS N, however, the presence of $\Delta$ baryons significantly changes $M$ and $R_e$, specially when the star is hot. If $\Delta$ resonances are present in the stellar matter, the gravitational mass decreases more quickly as the PNS deleptonizes. This is attributed to the softening of the EoS as new degrees of freedom are added to the stellar matter, lowering thermal energy per baryon and decreasing the core temperature as shown in \Cref{figrext}. At the same time, if we assume the PNS with the EoS NHD are static or lose angular momentum as they evolve, their radius is always smaller than with the EoSs N and NH. On the other hand, in the scenario with conserved angular momentum, the PNS with the NHD EoS has a higher radius at the stage $s_B=2, \,Y_{\nu_e}=0$. This occurs because at this specific configuration the PNS reaches the Keplerian frequency, which leads the star to have its highest equatorial radius possible, as shown in the top left panel of \Cref{fignB2}. Although the appearance of $\Delta$-resonances softens the EoS and would typically promote contraction, conservation of angular momentum instead forces the star to expand to avoid excessive spin-up. This constraint leads to the significant radius increase observed, unlike the slight contractions in the other two scenarios. Furthermore, by the third stage of evolution, neutrinos have largely escaped from the stellar core, removing their pressure support. This absence makes the star more susceptible to rotational effects, thereby amplifying the impact of rotation during this phase \cite{Spruit:1999cc, Prakash:1996xs}.

On the right side of \Cref{figpvst}, we show the changes in moment of inertia and frequency of a PNS with $M_0=1.60$~\msun~as it evolves. We can observe that in the scenario with conserved angular momentum, the moment of inertia increases as the PNS delepnotizes, which is associated with the matter redistribution inside the star caused by the increase in the temperature and entropy. On the other hand, in the scenario with loss of angular momentum, the value of $I$ slightly decreases as the PNS deleptonizes, mainly because the moment of inertia is directly proportional to the angular momentum. As for the frequency of rotation of the star, it decreases in all scenarios when the entropy increases from $s_B=1$ to $s_B=2$, which is justified by the increase in temperature and decrease in central density and compactness. However, in the next step of evolution, when the PNS goes from $s_B=2, \,Y_L=0.2$ to $s_B=2, \,Y_{\nu_e}=0$, the variations in $T$, $n_c$ and $M/R_e$ are smaller and the value of $\Omega$ can increase or decrease depending on the composition and the scenario considered. In both scenarios $(\dot{J}=0 \text{ and } \dot{J}<0)$, the moment of inertia and the rotational frequency increase when the star becomes a cold NS, since at this stage the star reaches its highest compactness. We can also observe that the EoSs for N and NH exhibit similar behavior for the variations in $I$ and $\Omega$, while the EoS with $\Delta$s is the one that produces more different results. In addition, the EoS NHD is the one that shows greater differences when comparing the scenario with and without conservation of the angular momentum. 
\begin{figure}[ht]
  \centering
    \includegraphics[width=0.75\linewidth]{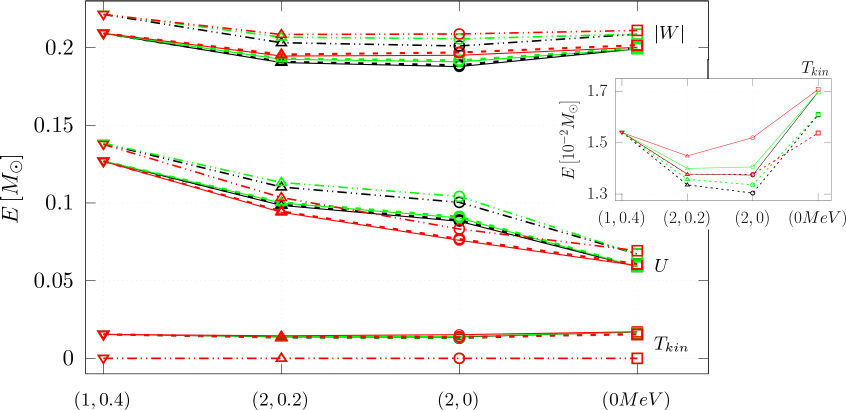}
  \caption{Evolution of the rotational energy $({T_{\rm kin}})$, internal energy $(U)$ and absolute value of the gravitational binding energy $(|W|)$ for a NS with $M_0 = 1.60$~\msun. Solid lines are for conserved angular momentum, dashed lines are for loss of angular momentum by neutrino emission, and dash-double-dot lines are for static stars. Inverted triangles are for $s_B=1, \,Y_L=0.4$, triangles are for $s_B=2, \,Y_L=0.2$, circles are for $s_B=2, \,Y_{\nu_e}=0$ and squares are for $T=0$ MeV.}
    \label{figevst}
  \label{fig:main_with_inset}
\end{figure}

\Cref{figevst} illustrates the evolution of various stellar energies, the rotational kinetic energy ${T_{\rm kin}}$, internal energy $U$, and gravitational binding energy $|W|$, for the evolution of a PNS with a baryonic mass of $M_0 = 1.60$~\msun. The ${T_{\rm kin}}$ plays a crucial role in supporting the star against gravity \cite{Lattimer:2000nx, Glendenning2000}, higher ${T_{\rm kin}}$ provides greater centrifugal support, leading to an increased equatorial radius, while lower ${T_{\rm kin}}$ allows the star to contract \cite{friedman1988ApJ, 1994ApJ...424..823C}. In the static case, represented by dash-double-dot lines, ${T_{\rm kin}}$ remains constant at zero. As the star evolves, higher entropy states tend to correlate with a decrease in the $T_{\rm kin}$, since elevated temperatures and larger radii decrease the spin of the PNS. Among the three scenarios analyzed, the conserved angular momentum case consistently exhibits higher ${T_{\rm kin}}$ than the case where angular momentum is lost through neutrino emission \cite{Duncan1986}, since in this last case the decrease in spin is more pronounced. Additionally, introducing more degrees of freedom in the stellar matter, for instance, the presence of $\Delta$-resonances tends to increase ${T_{\rm kin}}$. The $\Delta$-resonances slightly raise ${T_{\rm kin}}$ compared to hyperonic matter, which itself yields higher ${T_{\rm kin}}$ than purely nucleonic matter, due to their influence on the EoS and compacting the stellar structure. By inference, a softer EoS is associated with higher ${T_{\rm kin}}$ \cite{Glendenning2000}. The inset panel focuses on the region $1.3 \leq T_{\rm kin} \, [10^{-2} \, M_{\odot}] \leq 1.7$ to highlight the clear distinction in $T_{\rm kin}$ based on angular momentum and stellar composition differences, as discussed above. From the inset, we deduce that $T_{\rm kin}$ increases with stellar compactness. Its value decreases from the neutrino-trapping stage to the deleptonization phase, then increases slightly during the neutrino-transparent stage in the scenario with $\dot{J}=0$, while it decreases slightly in the scenario with $\dot{J}<0$. Then, the kinetic energy reaches its maximum when the star becomes cold and catalyzed. This trend follows the direction of increasing $\Omega$, which is directly proportional to $T_{\rm kin}$.

The internal energy $U$ of a PNS arises from thermal motion, Fermi degeneracy pressure, and nuclear interactions~\cite{Prakash:1996xs, Haensel:2007yy}. During deleptonization, $U$ decreases as the star cools and neutrinos escape, marking a transition toward a cold, catalyzed configuration \cite{Pons:1998mm}. The emergence of additional degrees of freedom, such as hyperons and $\Delta$-resonances, initially increases $U$ due to their mass and entropy contributions. However, by softening the EoS, these particles lower pressure support, promote contraction, and ultimately reduce the internal energy per baryon \cite{Drago:2014oja, Weissenborn:2011ut}. Interestingly, in the static case, the NH composition shows a slightly higher $U$ than stars composed purely of nucleons, reflecting the complex interplay between composition and gravitational binding. Rotational dynamics further influence $U$: static configurations exhibit the highest $U$ due to tighter gravitational binding, while models conserving angular momentum maintain lower $U$ as rotational support redistributes energy and reduces central compression, consistent with virial equilibrium and the interplay between gravitational, thermal, and rotational energy components.

The magnitude of $|W|$ reflects the strength of the star's gravitational potential, a higher $|W|$ corresponds to a more tightly bound and compact configuration \cite{Lattimer:2004pg, Shapiro:1983du}. In the first stage of evolution, when $s_B = 1$ and $Y_l = 0.4$, the star is relatively cooler and more compact, resulting in a higher $|W|$ compared to the intermediate stages. As the entropy increases to $s_B = 2$ and the lepton fraction decreases, the star becomes thermally supported and expands, leading to a lower central density and a temporary reduction in $|W|$ \cite{1986ApJ...307..178B}. During the transition from the deleptonization phase to the neutrino-transparent stage, $|W|$ remains relatively stable. As the star continues to cool and contracts into a cold, catalyzed configuration, $|W|$ increases again, reflecting the growing compactness of the system \cite{Glendenning2000}. Furthermore, the presence of additional degrees of freedom in the stellar matter, such as hyperons or $\Delta$-resonances, leads to an increase in $|W|$. These new particle species raise the total mass-energy content and soften the EoS, which promotes further contraction. The resulting increase in compactness enhances the gravitational binding energy, thus increasing~$|W|$~\cite{Bombaci:2016xzl}.

\section{Evolving star under the PSR J0740$+$6620 constraint} \label{sec_nicer}

\begin{table}[!ht]
 \caption{Rotational and compositional properties of evolving massive PNS.}
    \centering
    \setlength\extrarowheight{2pt}
    \begin{ruledtabular}
    \begin{tabular}{ c c c c c c c c c c c c c c }
   $s_B,\,Y_L$ & EoS &\multicolumn{4}{c}{$J$ conserved}&
\multicolumn{4}{c}{$J$ loss}&\multicolumn{4}{c}{static}\\
  \cline{3-6} \cline{7-10} \cline{11-14}
      &   & $J[M_{\odot}^2]$ & $n_c/n_0$ & $f_R[\%]$ & $M/R_e$ & $J[M_{\odot}^2]$ & $n_c/n_0$ & $f_R[\%]$ & $M/R_e$ & $J[M_{\odot}^2]$ & $n_c/n_0$ & $f_R[\%]$ & $M/R_e$ \\
        \hline
                 & \ft N & 0.76 & 3.14 & 2.5 & 0.23 & 0.78 & 3.14 & 2.5 & 0.23 & 0 & 3.20 & 0 & 0.23 \\

        $1,\, 0.4$ & \ft NH & 0.74 & 3.37 & 2.5 & 0.23 & 0.76 & 3.37 & 2.5 & 0.23 & 0 & 3.46 & 0 & 0.24 \\

                 & \ft NHD & 0.72 & 3.35 & 2.5 & 0.23  & 0.78 & 3.34 & 2.5 & 0.23 & 0 & 3.43 & 0 & 0.24 \\
        \hline 
                 & \ft N & 0.76 & 2.83 & 2.5 & 0.22 & 0.77 & 2.83 & 2.5 & 0.22 & 0 & 2.88 & 0 & 0.22 \\

        $2, \,0.2$ & \ft NH & 0.74 & 3.37 & 2.5 & 0.22 & 0.75 & 3.37 & 2.5 & 0.22 & 0 & 3.46 & 0 & 0.23 \\

                 & \ft NHD & 0.72 & 3.30 & 2.5 & 0.22 & 0.75 & 3.30 & 2.5 & 0.22 & 0 & 3.37 & 0 & 0.23 \\
        \hline
                & \ft N & 0.76 & 2.82 & 2.5 & 0.21 & 0.76 & 2.82 & 2.5 & 0.21 & 0 & 2.87 & 0 & 0.22 \\

        $2,\, 0$  & \ft NH & 0.74 & 3.47 & 2.5 & 0.22 & 0.74 & 3.47 & 2.5 & 0.22 & 0 & 3.58 & 0 & 0.23 \\

                & \ft NHD & 0.72 & 3.38 & 2.5 & 0.22 & 0.72 & 3.38 & 2.5 & 0.22 & 0 & 3.45 & 0 & 0.23 \\
        \hline
                & \ft N & 0.76 & 3.02 & 3.3 & 0.23 & 0.76 & 3.02 & 3.3 & 0.23 & 0 & 3.07 & 0 & 0.23 \\
                
        $T = 0$ & \ft NH & 0.74 & 3.43 & 2.5 & 0.23 & 0.74 & 3.43 & 2.5 & 0.23 & 0 & 3.53 & 0 & 0.24 \\
        
                & \ft NHD & 0.72 & 3.50 & 2.5 & 0.24 & 0.72 & 3.50 & 2.5 & 0.24 & 0 & 3.59 & 0 & 0.24 \\                
    \end{tabular}
    \end{ruledtabular}
    \label{tab2}
\end{table}
In this section, we investigate the properties of a PNS that, in the last stage, when the star is already a cold catalyzed NS, has the same gravitational mass and rotational rate of PSR J0740$+$6620, that is, $M=2.08$~\msun~and $\Omega=346.53$ Hz~\cite{NANOGrav:2019jur, Fonseca:2021wxt}. In this approach, we examine the evolutionary stages of the PNS,  
beginning with its final stage ($T=0$ MeV), and going backward to infer the properties it must have possessed during earlier stages to ultimately reproduce the observed mass and 
angular momentum of PSR J0740$+$6620. We start by calculating the baryonic mass $(M_0)$ and angular momentum of a NS ($T=0$ MeV) with the same gravitational mass and rotational rate as PSR J0740$+$6620 for each of the EoSs considered. For the EoSs N and NH, we obtain $M_0=2.39$~\msun~and $J=0.76$ and $0.74\, M_{\odot}^2$, respectively, and for the NHD EoS we obtain $M_0=2.40$~\msun~and $J=0.72\, M_{\odot}^2$. For the static scenario, we found that a NS with a baryonic mass of $M_0=2.39$~\msun~has a gravitational mass of $M=2.08$~\msun~for all three EoSs. For the static scenario, we calculate the properties of the previous stages of evolution keeping the value of $M_0$ fixed, while for the scenario of conserved angular momentum, we keep $M_0$ and $J$ fixed with the same values they have in the stage with $T=0$ MeV for each EoS. For the scenario with loss of angular momentum, we invert Eq. \eqref{eq1} in the following way:
\begin{equation}
    J_i=J_f\left( \frac{M_{i}}{M_{f}}\right)^{2.55}, \label{eq2}
\end{equation}
since now we know the values of $J_f$ and want to find $J_i$. 

In \Cref{tab2}, we present the evolution of the angular momentum $J$, central baryonic density $n_c/n_0$, deformation $f$, and compactness $M/R_e$ for each of the three scenarios and three EoSs. The PNS analyzed in this section rotates at a frequency approximately half that of the PNS discussed in the previous section, and consequently, its angular momentum is also roughly half the value shown in \Cref{tab1}. We find that when the stellar matter contains only nucleons, a higher angular momentum is required to achieve the same rotational frequency as in the cases where hyperons are also present, when $\Delta$ baryons are included, $J$ is even smaller. Conversely, as more degrees of freedom are introduced, higher central baryonic densities are required to produce a NS with the same gravitational mass ($M = 2.08$~\msun).

As expected, $n_c/n_0$ values are higher for the star analyzed in this section than in the previous one, since its mass is greater. Notably, although the rotation frequency is now about half its previous value, the star’s deformation is roughly ten times smaller. This occurs because the star is both rotating more slowly and more massive, making it more resistant to deformation, as confirmed by its compactness, which is now about twice as high. From the data in \Cref{tab2}, we can infer that PSR J0740$+$6620 would be more deformed if composed solely of nucleons than if its core also contained hyperons or hyperons plus $\Delta$ baryons.

Returning to the early stage with entropy per baryon $s_B = 2$ and no electron neutrinos ($Y_{\nu_e} = 0$), the central baryon density decreases for the N and NHD EoSs, likely due to thermal expansion. However, for the NH EoS, the central density is unexpectedly higher in this hot stage than in the cold stage ($T = 0$ MeV). Looking further back, during the neutrino-opaque stage $(s_B = 2\, , Y_L = 0.2)$, the NH and NHD EoSs yield a lower central density than in the later neutrino-transparent stage $(s_B = 2\, , Y_{\nu_e} = 0)$. In contrast, for the N EoS, the central density is higher when the star is still neutrino-opaque. In the earliest stage considered $(s_B = 1\, , Y_L = 0.4)$, corresponding to the just-born PNS, the central baryon density is highest for all EoSs, indicating that the star is most compact at birth.

When comparing the three EoSs under angular momentum loss with those under angular momentum conservation, we find that for the N and NH compositions, there is an overall decrease of $2.6\%$ in angular momentum, while for the NHD EoS, the decrease amounts to $7.7\%$. All angular momenta were calculated under the same physical conditions for consistency. Among the properties listed in \Cref{tab2}, only the central baryon density for the NHD EoS at stage $(s_B = 1, \,Y_L = 0.4)$ shows a slight sensitivity to angular momentum loss, with a small change of just $0.3\%$. As for the static configuration, it leads to stars that are more compact and have higher central baryon density than when rotation is taken into account.
\begin{figure}[t!]
    \centering
    \includegraphics[width=0.49\linewidth]{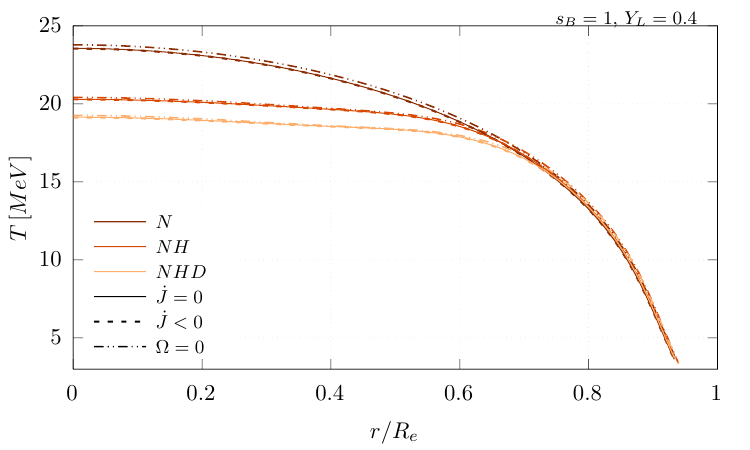}
    \includegraphics[width=0.49\linewidth]{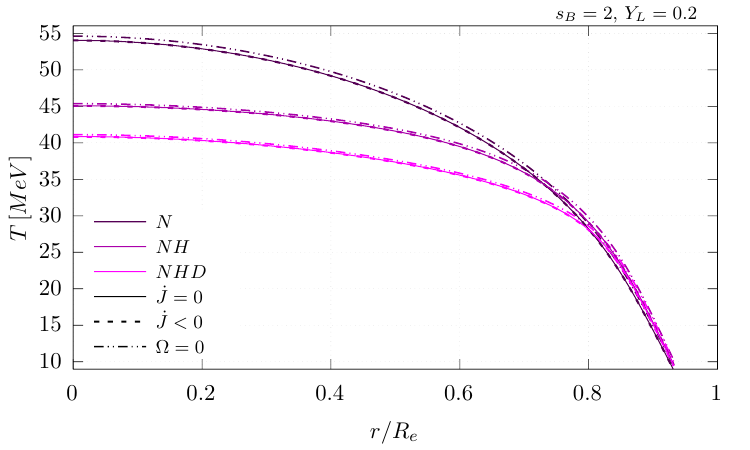}
    \includegraphics[width=0.49\linewidth]{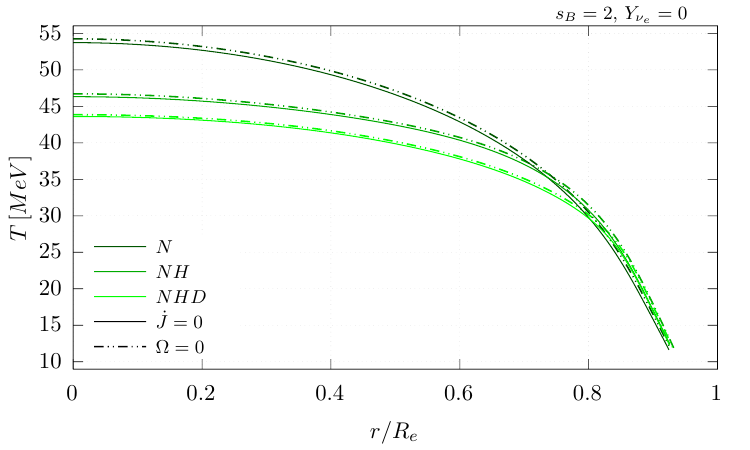}  
    \caption{Profiles of temperature versus normalized equatorial radius for a star with $M = 2.08$~\msun~at the $T=0$ MeV stage, solid lines are for conserved angular momentum, dashed lines are for loss of angular momentum by neutrino emission, and dash-double-dot lines are for static stars. {Top left: $s_B=1,\,Y_L=0.4$, top right: $s_B=2,\,Y_L=0.2$, bottom: $s_B=2,\,Y_{\nu_e}=0$. }}
    \label{figrext2}
\end{figure}

\Cref{figrext2} shows the temperature profiles as a function of normalized equatorial radius for an evolving PNS whose final state corresponds to PSR J0740$+$6620. The qualitative trends mirror those in \Cref{figrext}, but this case incorporates a more realistic scenario based on an observed, slowly rotating pulsar. Our modeling ensures that the final cold state aligns with observational constraints, despite the absence of direct PNS data. Due to the higher mass considered here, the PNS exhibits systematically elevated temperatures at all three evolutionary stages compared to the reference case in \Cref{figrext}. However, the relatively low rotational frequency results in minimal departures from spherical symmetry, and angular momentum loss introduces no significant variation in the thermal profiles. A particularly notable feature arises for the purely nucleonic composition: during the deleptonization stage $(s_B = 2,\, Y_L = 0.2)$, the PNS core is slightly hotter than in the neutrino-transparent stage $(s_B = 2,\, Y_{\nu_e} = 0)$. This behavior is linked to a modest increase in central baryon density at $Y_L = 0.2$, as shown in \Cref{tab2}, reflecting the additional pressure support provided by trapped neutrinos. This observation implies that, in massive, slowly rotating PNSs like the ones that lead to the formation of PSR J0740$+$6620, trapped neutrinos enhance core compression and heating during deleptonization, while the slow rotation renders the thermal evolution largely insensitive to angular momentum loss, a key constraint for modeling realistic PNS cooling scenarios.
\begin{figure}[t!]
    \centering
    \includegraphics[width=0.49\linewidth]{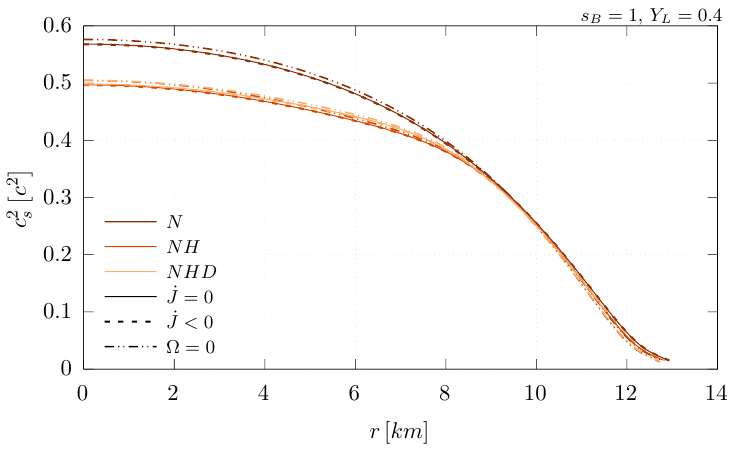}
    \includegraphics[width=0.49\linewidth]{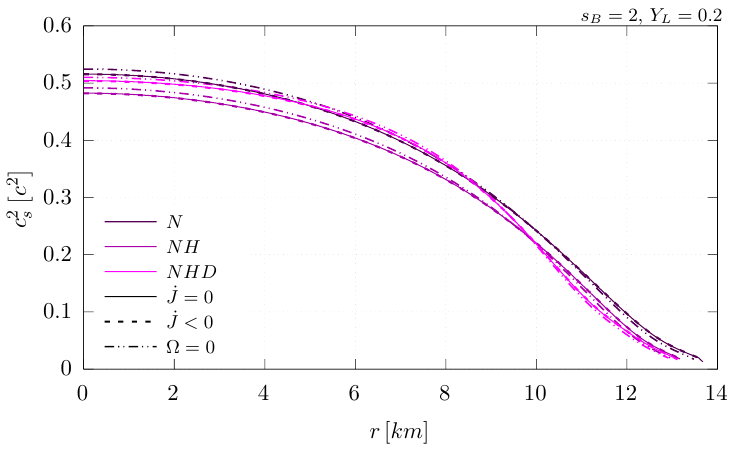}
    \includegraphics[width=0.49\linewidth]{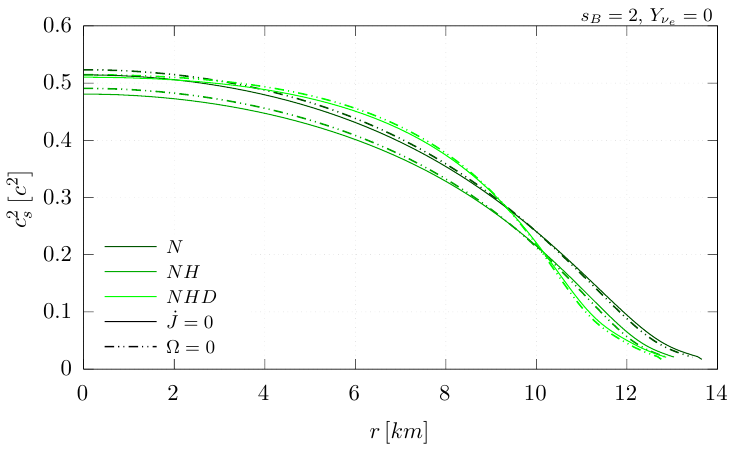}
    \includegraphics[width=0.49\linewidth]{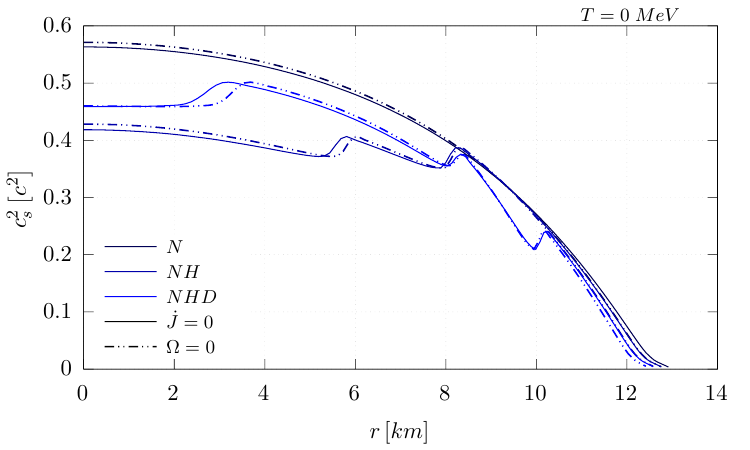}
    \caption{Profiles of squared sound velocity versus the equatorial radius for a star with $M = 2.08$~\msun~at the $T=0$ MeV stage, solid lines are for conserved angular momentum, dashed lines are for loss of angular momentum by neutrino emission and dash-double-dot lines are for static stars.}
    \label{figrexcs2c}
\end{figure}

\Cref{figrexcs2c} exhibits similar qualitative trend as \Cref{figrexcs}, in terms of thermodynamic behavior and particle composition but incorporates realistic rotational dynamics based on the observed properties of PSR J0740$+$6620, a slowly rotating ($\Omega=346.53$ Hz), massive ($M=2.08$~\msun) pulsar with a NICER-constrained radius of $\approx 12.39-13.70$ km \cite{riley2021,miller2021}. While both figures display comparable evolutionary patterns, \Cref{figrexcs2c} shows less pronounced rotational splitting in the $c_s^2$ profiles due to the pulsar's slower spin. At the same time, its high mass necessitates a stiffer EoS, leading to systematically elevated core sound speeds. This comparison demonstrates how observed pulsar properties, mass, radius, and spin, quantitatively reshape PNS thermodynamics, even as the overall qualitative behavior remains consistent with more idealized scenarios like those in \Cref{figrexcs}.
\begin{figure}[t!]
    \centering
    \includegraphics[width=0.49\linewidth]{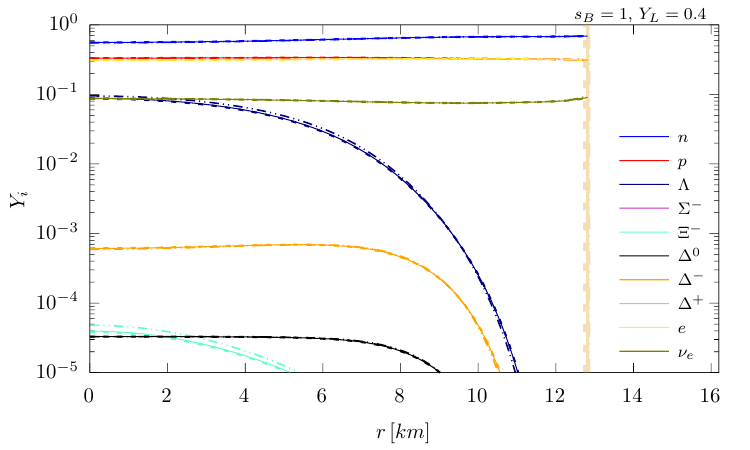}
    \includegraphics[width=0.49\linewidth]{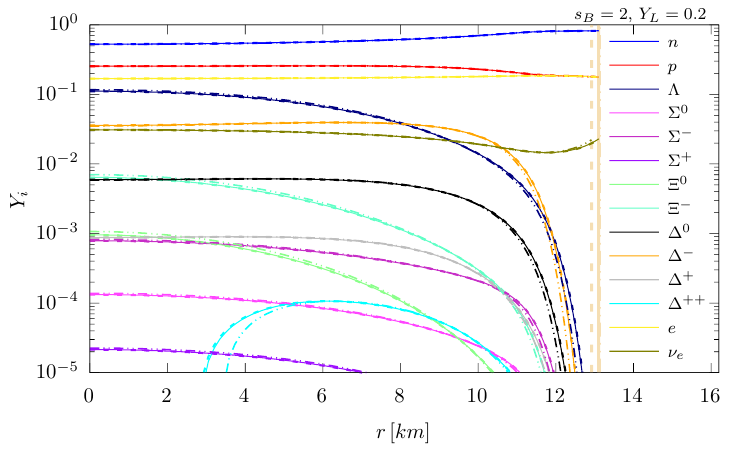}
    \includegraphics[width=0.49\linewidth]{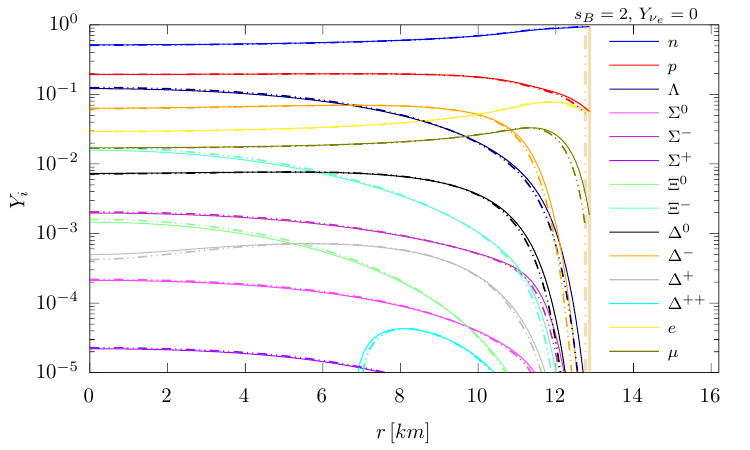}
    \includegraphics[width=0.49\linewidth]{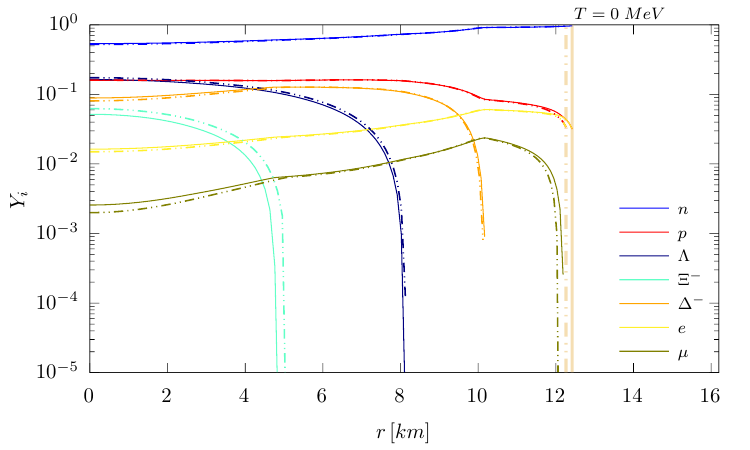} 
    \caption{Profiles of particle composition versus the equatorial radius for a star with $M = 2.08$~\msun~at the $T=0$ MeV stage, solid lines are for conserved angular momentum, dashed lines are for loss of angular momentum by neutrino emission, and dash-double-dot lines are for static stars. Vertical lines represent the respective radius were the stellar crust begins.}
    \label{figrexyi2}
\end{figure}

\Cref{figrexyi2} shows the particle distribution in the massive, slowly rotating pulsar PSR J0740$+$6620, which can be compared with the rapidly rotating case in \Cref{figrexyi}, corresponding to a star with $M_0 = 1.60$~\msun. While both exhibit qualitatively similar trends, PSR J0740$+$6620 displays notable differences due to its greater mass and slower rotation. In the neutrino-trapped stage, additional hyperons ($\Sigma^-$ and $\Xi^-$) and the $\Delta^+$ resonance appear in PSR J0740$+$6620, in contrast to the $M_0=1.60$~\msun~model which shows only $\Lambda$ hyperons and $\Delta^0$, $\Delta^-$ states. Shared species such as $\Lambda$, $\Delta^0$, and $\Delta^-$ appear at larger radii in PSR J0740$+$6620, closer to the surface, indicating lower onset densities (e.g., $\Lambda$ at $\sim11$ km vs. $\sim9.7$ km, $\Delta^-$ at $\sim10.5$ km vs. $\sim8$ km under $J = \mathrm{const}$), with $\Lambda$ also being more abundant. This trend continues through the deleptonization and neutrino-transparent stages, where higher temperatures enhance thermal production, leading to substantial populations of all hyperons and $\Delta$-resonances, again shifted outward in the massive star. In the final cold, catalyzed stage, PSR J0740$+$6620 retains the $\Delta^-$ resonance and the hyperons $\Xi^-$ and $\Lambda$, whereas the less massive model shows only the hyperon $\Lambda$. Across all evolutionary phases, particle abundances follow a rotational hierarchy, most abundant and earliest in the static configuration, followed by the $\dot{J}<0$ case, and least in the $J = \mathrm{const}$ scenario. These results emphasize the role of stellar mass and rotation in shaping particle composition and its spatial distribution throughout the PNS evolution.

\begin{figure}[t!]
    \centering
    \includegraphics[width=0.49\linewidth]{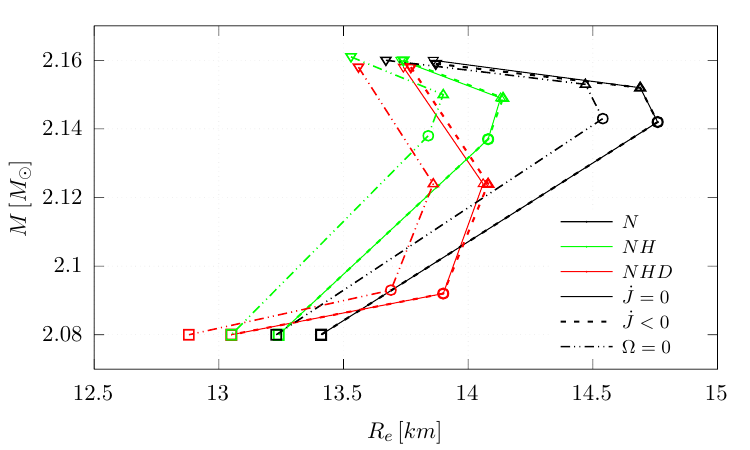}
    \includegraphics[width=0.49\linewidth]{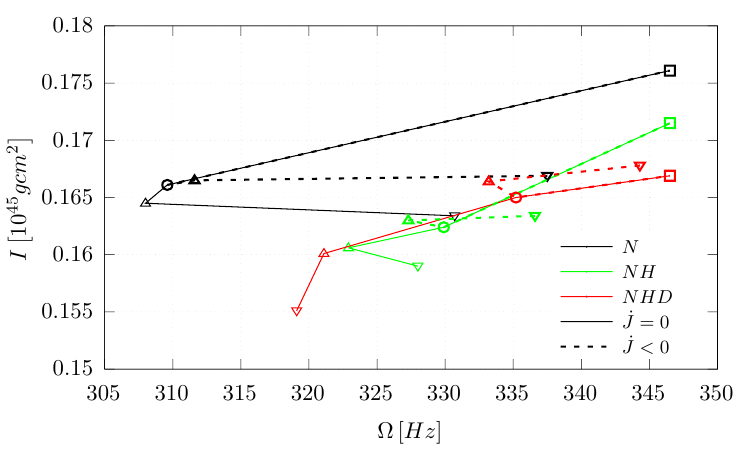}
    \caption{Evolution of the gravitational mass and equatorial radius, on the left, and of the moment of inertia and angular frequency, on the right, for a NS with $M = 2.08$~\msun~at the $T=0$ MeV stage. Solid lines are for conserved angular momentum, dashed lines are for loss of angular momentum by neutrino emission and dash-double-dot lines are for static stars. Inverted triangles are for $s_B=1, \,Y_L=0.4$, triangles are for $s_B=2, \,Y_L=0.2$, circles are for $s_B=2, \,Y_{\nu_e}=0$ and squares are for $T=0$ MeV.}
    \label{figpvst2}
\end{figure}

The left panel of \Cref{figpvst2} shows the evolution of the gravitational mass and equatorial radius of PSR J0740$+$6620 for each scenario and composition considered in this work. As in the previous section, the gravitational mass decreases as the PNS evolves, with the inclusion of additional degrees of freedom causing the mass to drop more rapidly during deleptonization. The equatorial radius increases during the hotter stages and then decreases again as the star cools. Since a smaller rotation rate is considered here, the differences between the static and rotating cases are less pronounced than for the $M_0 = 1.60$\msun~analyzed previously. The scenario with $\dot{J} < 0$ deviates only slightly from the $\dot{J} = 0$ case, and only for the NHD EoS. For all cases studied, the predicted equatorial radii are consistent with the results of Riley et al.\cite{riley2021} and Miller et al.\cite{miller2021}. When rotation is included, all three EoSs yield values in the range $13 < R_e \, [\mathrm{km}] < 13.5$.

The right panel of \Cref{figpvst2} shows the evolution of the moment of inertia and rotational frequency as the PNS evolves. In the final stage, all cases reach the same frequency as PSR J0740$+$6620 ($346$ Hz). The moment of inertia required to achieve this frequency decreases as the number of degrees of freedom increases, since the inclusion of hyperons and $\Delta$ baryons in the EoS makes the star more compact. As in the previous case, for all EoSs in the $\dot{J} = 0$ scenario, the moment of inertia increases throughout the evolution, while in the $\dot{J} < 0$ scenario it decreases during deleptonization and then rises again once the star cools into a NS. The rotational frequency trends are also similar to those in the previous case, except that for the NHD EoS in the $\dot{J} = 0$ scenario, $\Omega$ increases steadily throughout the evolution. Once again, the NHD EoS exhibits the most pronounced variations in both $I$ and $\Omega$ when comparing the cases with and without angular momentum loss. For this EoS, the value of $\Omega$ in the first stage differs by $7.9\%$ between the two scenarios.

\begin{figure}[t!]
    \centering
    \includegraphics[width=0.75\linewidth]{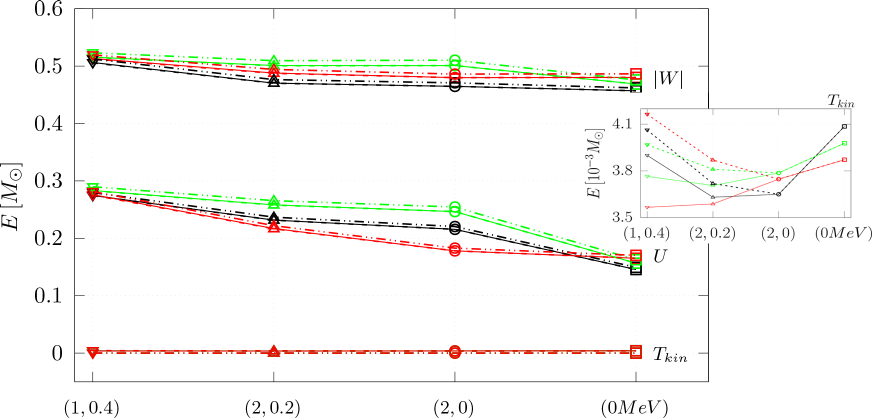}
    \caption{Evolution of the rotational kinetic energy $(T_{\rm kin})$, internal energy $(U)$ and absolute value of the gravitational binding energy $(|W|)$ for a NS with $M = 2.08$~\msun~at the $T=0$ MeV stage. Solid lines are for conserved angular momentum, dashed lines are for loss of angular momentum by neutrino emission, and dash-double-dot lines are for static stars. Inverted triangles are for $s_B=1, \,Y_L=0.4$, triangles are for $s_B=2, \,Y_L=0.2$, circles are for $s_B=2, \,Y_{\nu_e}=0$ and squares are for $T=0$ MeV.}
    \label{figevst2b}
\end{figure}

\Cref{figevst2b} shows the energy evolution of PSR J0740$+$6620 from birth to its cold state. The evolution of $T_{\rm kin}$ reflects a complex interplay between stellar mass, rotational energy, and composition, distinct from the hypothetical $M_0=1.60$\msun~shown in \Cref{figevst}. The inset, focusing on $3.5 \leq T_{\rm kin} \, [10^{-3} \, M_{\odot}] \leq 4.1$, shows that NHD starts with the lowest $T_{\rm kin}$ in the scenario with conserved angular momentum, followed by NH and N. However, when angular momentum loss is included, the NHD EoS exhibits the highest $T_{\rm kin}$ in the first stage, reflecting the increased sensitivity of stars with more degrees of freedom to rotational dynamics. During deleptonization, $T_{\rm kin}$ for N drops rapidly, while NH decreases more gradually. For the NHD EoS, kinetic energy decreases if angular momentum is lost but increases if $J$ remains constant. At the neutrino-transparent stage, $T_{\rm kin}$ for N becomes the lowest, followed by NHD, while NH remains the highest. Toward the cold and catalyzed configuration, all compositions show an increase in $T_{\rm kin}$, however, N rises more rapidly, overtaking NH and ending with the highest value, followed by NH, then NHD. This inversion contrasts with \Cref{figevst}, where $T_{\rm kin}$ decreases with increasing degrees of freedom. It is justified by the fact that, when angular momentum is lost, a higher kinetic energy is required to achieve the same final rotation frequency. 

 For the $\dot{J}<0$ case, the $T_{\rm kin}$ of NHD dominates during the neutrino-trapping phase, followed by N and NH. This dominance persists through deleptonization, with NH as the second most abundant and N the least. From the neutrino-transparent stage to the formation of a cold, catalyzed configuration, the evolution of $T_{\rm kin}$ follows the same trend as in the $\dot{J}=0$ case. For the star considered in this section, the values of $U$ and $|W|$ are higher than for the star in the previous section due to its larger mass, while $T_{\rm kin}$ remains small because of the lower rotation rate. In both stars, $T_{\rm kin}$ vanishes in static configurations. This highlights that mass and rotation, rather than composition, primarily govern the thermodynamic evolution.

Throughout all stages, from neutrino-trapped to the final cold state, $U$ decreases due to the loss of thermal and neutrino pressure support. Notably, the composition dependence of $U$ is non-monotonic: NH matter consistently exhibits the highest $U$, followed by N matter, with NHD matter showing the lowest values. This inverted ordering persists during the deleptonization and neutrino-transparent phases. In the final cold, catalyzed state, however, the expected trend is restored, with $U$ increasing with the number of degrees of freedom. This highlights the complex interplay of mass, rotation, and composition in shaping the thermodynamic evolution of a PNS. The $|W|$ deviates from the hypothetical trend. Early on, N stars show the lowest $|W|$, NHD intermediate, and NH the highest. In the cold phase, the expected increase of $|W|$ with compositional complexity is recovered. Importantly, NH and NHD exhibit a U-shaped $|W|$ evolution, a drop during deleptonization, followed by a rise due to contraction from neutrino loss. This is absent in N matter, which shows a more flat trend. 

\section{Final Remarks and Conclusion} \label{remarks}

We investigated the local properties of rotating PNSs, including temperature profiles, speed of sound, and particle composition, and how these microphysical features connect to global observables such as gravitational mass, equatorial radius, and moment of inertia. This multi-scale approach aims to bridge microscopic physics with macroscopic astrophysical signatures relevant for current and future multimessenger constraints. We adopt two complementary strategies. First, we model an hypothetical PNS with baryon mass $M_0 = 1.60$~\msun~(corresponding to a gravitational mass of $\sim 1.4$~\msun~at $T=0$ MeV) and rotating at a frequency close to the Kepler limit under three conditions: angular momentum conservation ($J = \mathrm{const}$), angular momentum loss ($\dot{J} < 0$), and the static case ($\Omega = 0$). Second, we simulate the realistic slowly rotating pulsar PSR J0740$+$6620, whose mass and radius are precisely constrained by NICER observation. We enforce that the final cold and catalyzed configuration of the star must be consistent with PSR J0740$+$6620's observed gravitational mass and rotational frequency \cite{NANOGrav:2019jur, Fonseca:2021wxt}. Additionally, our EoSs lead to an equatorial radius that agrees with the NICER constraints \cite{riley2021,miller2021}. Comparing its internal composition, thermal evolution, and energy evolution to the rapidly rotating hypothetical case near the Keplerian limit to quantify how rotation alters microphysical structure, energy transport, and the mass evolution of the star.

The results in \Cref{tab1} show that stellar deformation is dynamically regulated by the combined effects of angular momentum, thermal pressure, and the EoS of dense matter \cite{1994ApJ...424..823C}. When $J = \mathrm{const}$, the star's deformation increases due to centrifugal forces, especially during the neutrino-transparent phase, when thermal pressure peaks \cite{Pons:1998mm} and the EoS softens \cite{Malfatti:2020onm}. The corresponding rise in $n_c$ observed in the $\dot{J} < 0$ scenario indicates angular momentum redistribution leading to stellar contraction \cite{camelio2016, Martinon:2014uua}. This scenario also presents higher deformation when the star is hotter (the intermediate stages) due to high entropy, similar to the $J = \mathrm{const}$ case \cite{Farrell:2024rfa}. The maximal deformation $51\%$ ($J = \mathrm{const}$, scenario) observed occurred when the star's particle degrees of freedom are higher, for instance, through the presence of $\Delta$-resonances. This underscores how exotic particles and finite-temperature effects collectively shape rotational instabilities by softening the EoS. Conversely, a cold and catalyzed matter, characterized by reduced thermal pressure and a stiffer EoS, resists deformation \cite{Glendenning2000}. A similar observation was made in \Cref{tab2}, where we consider a more realistic case based on the known slow rotation data for PSR J0740$+$6620. In this case, the changes induced by rotation are minimal, with stellar deformation remaining well below approximately $3.3\%$ \cite{riley2021} across all the stages analyzed.  These findings suggest that PNSs formed in core-collapse supernova or binary mergers could undergo significant transient deformations, potentially influencing pulsar kicks. 

The detailed outcomes of our analysis are presented below:
\begin{itemize}
    \item The temperature profiles of a $1.60$~\msun~PNS depend strongly on composition, rotation, and evolutionary stage. Nucleonic matter maintains the highest temperatures, while the presence of hyperons and $\Delta$-resonances lowers them by distributing thermal energy across more degrees of freedom. Rotating configurations with $J = \text{const}$ are associated with smaller core temperatures due to centrifugal support and reduced $n_c$, whereas static stars reach higher densities and temperatures compared to $J = \text{const}$ and $\dot{J} < 0$ cases. As neutrino pressure fades, temperature differences become more pronounced, especially during the neutrino-transparent phase where the temperature of the stellar matter peaks. The equatorial temperature gradient, $|\nabla T| \propto R_e^{-1}$, reflects the redistribution of heat due to rotational flattening. These results demonstrate how microphysics and rotation shape the thermal and structural evolution of PNSs. Notably, in the massive, slowly rotating pulsar PSR J0740$+$6620, thermal evolution is largely unaffected by angular momentum loss due to neutrino emission during the Kelvin-Helmholtz phase.
    
    \item The sound speed $c_s^2$ in rotating NSs reflects EoS stiffness: it decreases with the appearance of hyperons and $\Delta$-resonances \cite{Vidana:2015rsa, Issifu:2023qyi} and is further suppressed by centrifugal forces \cite{1994ApJ...424..823C}. As angular momentum is lost, $c_s^2$ tends toward static-star values, signaling core contraction, while the lowest value was observed in the $J = \text{const}$ due to higher centrifugal force and reduction in the $n_c$. In later stages, when the star is in cold and catalyzed configuration, bumps in the $c_s^2$ profile indicate thresholds for the appearance of new particle species in the stellar matter \cite{Glendenning2000}. Comparing a hypothetical $1.60$~\msun~PNS to the slowly rotating PSR J0740$+$6620 shows weaker rotational splitting in $c_s^2$, while preserving the microphysical features. This demonstrates that angular momentum loss alters global thermodynamics but retains compositional signatures linking the microscopic and macroscopic properties of the star. 

    \item The study reveals that particle abundances and thresholds in PNS are primarily governed by stellar mass, rotation, and thermal evolution. Higher mass, exemplified by PSR J0740$+$6620, enables particle diversity, introducing additional hyperons ($\Sigma^-$, $\Xi^-$ in the neutrino-trapped stage) and $\Delta$-resonances ($\Delta^+$), which appear earlier and at lower densities (larger radii) than in the less massive $M_0=1.60$~\msun~case considered. Rotation systematically suppresses particle abundances and delays the onset of hyperons and $\Delta$'s, following the hierarchy static $>$ ($\dot{J}>0$) $>$ $J = \rm const$ \cite{martinon2014PhRvD}, restricting core composition (e.g., only $\Delta^-$ in rotating cores vs. $\Delta^-$ and $\Delta^0$ in static for $s_B=1,\,Y_L=0.4$). Charge neutrality drives noticeable differences in the populations of $\Delta^+$ and $\Delta^{++}$ between static and rotating cases. In the final cold stage, only $\Delta^-$ persists, with rotation favoring neutrons, electrons, and muons over protons and hyperons. Thermal expansion throughout evolution significantly lowers particle thresholds (e.g., $\Delta^-$ from $1.64 \, n_0$ to $0.46 \, n_0$), enabling earlier appearances, while final cold stages show mass-dependent retention of exotic particles, the hypothetical case harbors only $\Lambda$ near the core, whereas the massive star retain both $\Lambda$ and $\Sigma^-$ \cite{Glendenning2000}. Our findings also reveal that higher temperatures lead to increased particle abundances and amplify stellar deformation \cite{Pons:1998mm, Issifu:2023qyi}, particularly at higher rotational frequencies.

    \item The energy evolution of both the hypothetical $1.60$~\msun~and PSR J0740$+$6620 cases highlights the roles of rotation, mass, and composition in stellar thermodynamics. In the hypothetical case, rotation increases radius and moment of inertia $I$, and $\Omega$, which raises $T_{\rm kin}$ and lowers compactness \cite{1994ApJ...424..823C}. Composition softens the EoS, deepening the gravitational well and increasing $|W|$ \cite{Malfatti:2020onm}. In case of PSR J0740$+$6620, slow rotation suppresses $T_{\rm kin}$ \cite{riley2021}, while high mass enhances both $U$ and $|W|$ \cite{Glendenning2000}. Unlike the hypothetical case, $U$ varies non-monotonically with composition, NH matter gives the highest $U$, followed by N and NHD for the PNS phase, though the standard ordering returns in the cold state. A U-shaped evolution of $|W|$ emerges in both stars, but it is more pronounced in the hypothetical case than in PSR J0740$+$6620. This results from competing effects: high compactness and low core temperature during the neutrino-trapped stage, reduced compactness but elevated core temperature during deleptonization, and increased compactness due to contraction in the cold, catalyzed stage. Thus, while the hypothetical case captures baseline behavior, realistic stars require EoS models that incorporate mass constraints, rotation, and composition-driven effects to explain observational features~\cite{Oertel:2016bki}.
\end{itemize}
Our findings demonstrate that the microscopic properties governing PNS evolution, composition, thermal gradients, energy evolution, and local thermodynamics are intrinsically linked to macroscopic observables such as birth mass, radius, rotation, and global structure. As a result, realistic modeling of PNSs demands an EoS constrained by mass, radius, and compositional data. An EoS constrained by such multiple observables is essential for precise evolutionary modeling and to improve the effectiveness of multimessenger astrophysical diagnostics.

\begin{acknowledgments}

A.I. acknowledges financial support from the São Paulo State Research Foundation (FAPESP), Grant No. 2023/09545-1. 
This work is part of the project INCT-FNA (Proc. No. 464898/2014-5) and is also supported by the National Council for Scientific and Technological Development (CNPq) under Grants Nos. 303490/2021-7 (D.P.M.) and 306834/2022-7 (T.F.) T. F. also thanks the financial support from  Improvement of Higher Education Personnel CAPES (Finance Code 001) and FAPESP Thematic Grants (2023/13749-1 and 2024/17816-8). 
L.C.N.S. would like to thank Conselho Nacional de Desenvolvimento Científico e Tecnológico - Brazil (CNPq) for financial support under Research Project No. 443769/2024-9 and Research Fellowship No. 314815/2025-2.
F.M.S. would like to thank CNPq for financial support under research project No. 403007/2024-0 and research fellowship No. 201145/2025-1.
\end{acknowledgments}

\bibliography{reference}

\end{document}